\documentclass[reprint,aps,prl,amsmath,amssymb,nofootinbib,superscriptaddress]{revtex4-1}

\usepackage{mathtools}
\usepackage{braket} 
\usepackage[dvipsnames]{xcolor}  
\usepackage{float}
\usepackage[T1]{fontenc}
\usepackage[utf8]{inputenc}
\usepackage{graphicx,color,rotating,pifont}
\usepackage{amsmath,amssymb}
\usepackage{mathrsfs}
\usepackage{mathtools}
\usepackage{subfigure}
\usepackage{siunitx}
\usepackage{bm}
\usepackage{ae}
\usepackage{dcolumn}
\usepackage{txfonts}
\usepackage{tensor}
\usepackage{braket}
\usepackage{booktabs}
\usepackage{xspace}
\usepackage{soul}
\usepackage{multirow}
\usepackage{tikz}
\usepackage{pgffor} 
\setstcolor{red}
\setul{}{.2ex} 

\usepackage{multirow} 
\usepackage{makecell}
 
\usepackage[normalem]{ulem}

\usetikzlibrary{backgrounds}
\usetikzlibrary{shapes,matrix,trees}
\usetikzlibrary{arrows.meta}
\usetikzlibrary{positioning}			
\usetikzlibrary{calc,through}			
\usetikzlibrary{positioning,calc,shapes,decorations.pathreplacing,decorations.markings,decorations.pathmorphing}
\tikzset{
	vector/.style={decorate, decoration={snake,amplitude=2.5pt}, draw},
	provector/.style={decorate, decoration={snake,amplitude=2.5pt}, draw},
	antivector/.style={decorate, decoration={snake,amplitude=-2.5pt}, draw},
	fermion/.style={draw=black, postaction={decorate},
		decoration={markings,mark=at position .6 with {\arrow[draw=black]{>}}}},
           vL/.style={draw=ppurple, postaction={decorate},
		decoration={markings,mark=at position .6 with {\arrow[draw=ppurple]{>}}}},
	vLp/.style={draw=ppurple, postaction={decorate},
		decoration={markings,mark=at position .7 with {\arrow[draw=ppurple]{>}}}},
	NR/.style={draw=ggreen, postaction={decorate},
		decoration={markings,mark=at position .62 with {\arrow[draw=ggreen]{>}}}},	
	NRp/.style={draw=ggreen, postaction={decorate},
		decoration={markings,mark=at position .7 with {\arrow[draw=ggreen]{>}}}},	
	neutralino/.style={draw=black},
	fermionbar/.style={draw=black, postaction={decorate},
		decoration={markings,mark=at position .6 with {\arrow[draw=black]{<}}}},
	fermionnoarrow/.style={draw=black},
	gluon/.style={decorate, draw=black,
		decoration={coil,amplitude=4pt, segment length=5pt}},
	scalar/.style={dashed,draw=black, postaction={decorate},
		decoration={markings,mark=at position .55 with {\arrow[draw=black]{>}}}},
	scalarbar/.style={dashed,draw=black, postaction={decorate},
		decoration={markings,mark=at position .55 with {\arrow[draw=black]{<}}}},
	scalarnoarrow/.style={dashed,draw=black},
	electron/.style={draw=black, postaction={decorate},
		decoration={markings,mark=at position .55 with {\arrow[draw=black]{>}}}},
	bigvector/.style={decorate, decoration={snake,amplitude=4pt}, draw},
	photon/.style={decorate, draw=black,decoration={snake,amplitude=4pt, segment length=5pt} }
}

\usepackage[colorlinks,linkcolor=blue,anchorcolor=blue,citecolor=blue]{hyperref} 
\hypersetup{
   colorlinks=true, 
   linkcolor=blue,  
   citecolor=blue,  
   filecolor=blue,  
   urlcolor=blue    
} 
 
\DeclareMathSymbol{\NS}{\mathord}{AMSb}{"4E}

\DeclareSIUnit{\fm}{\femto\meter}

\renewcommand{\vec}[1]{\ensuremath{\bm{#1}}}

\newcommand{\eMax}{\ensuremath{N_{\text{sh}}}}

\newcommand{\beq}{\begin{equation}}
\newcommand{\eeq}{\end{equation}}
\newcommand{\beqn}{\begin{eqnarray}}
\newcommand{\eeqn}{\end{eqnarray}}
\newcommand{\bsub}{\begin{subequations}}
\newcommand{\esub}{\end{subequations}}
\newcommand{\bpm}{\begin{pmatrix}}
\newcommand{\epm}{\end{pmatrix}}

\newcommand\identity{1\kern-0.25em\text{l}}

\makeatletter

\newcommand{\Rmnum}[1]{\expandafter\@slowromancap\romannumeral #1@}
\makeatother

\newcommand{\xx}[1]{\textcolor{red}{[xx]}}

\begin{document}
  
  \title{Impact of shape fluctuations on nuclear Schiff moments in heavy octupole-deformed nuclei }

 \author{E. F. Zhou}    
 \affiliation{School of Physics and Astronomy, Sun Yat-sen University, Zhuhai 519082, P.R. China} 
 
 \affiliation{Guangdong Provincial Key Laboratory of Quantum Metrology and Sensing, Sun Yat-Sen University, Zhuhai 519082, P. R. China }

  \author{J. M. Yao}   
 \email{Corresponding: yaojm8@sysu.edu.cn}
  \affiliation{School of Physics and Astronomy, Sun Yat-sen University, Zhuhai 519082, P.R. China}  
  
 \affiliation{Guangdong Provincial Key Laboratory of Quantum Metrology and Sensing, Sun Yat-Sen University, Zhuhai 519082, P. R. China }   

    \affiliation{Yukawa Institute for Theoretical Physics, Kyoto University, Kyoto, 606-8502, Japan}  
   
\author{J. Engel}     
\address{Department of Physics and Astronomy, University of North Carolina, Chapel Hill, North Carolina 27516-3255, USA}

\author{J. Meng}    
\address{State Key Laboratory of Nuclear Physics and Technology, School of Physics,  Peking University, Beijing 100871, China}
 
\date{\today}

\begin{abstract}
Permanent electric dipole moments (EDMs) are among the most sensitive probes of CP violation beyond the Standard Model. In diamagnetic atoms, the EDM is determined primarily by the nuclear Schiff moment, whose uncertainty limits the interpretation of current and future EDM searches. For the experimentally most promising octupole-deformed nuclei, however, existing Schiff-moment calculations have largely relied on the rigid-shape approximation. Here we report the first fully microscopic beyond-mean-field study of Schiff moments in the heavy octupole-deformed nuclei $^{225}$Ra, $^{229}$Th, and $^{229}$Pa, based on multireference covariant density functional theory (MR-CDFT) with symmetry restoration and quadrupole--octupole shape mixing. The results show that collective shape fluctuations reduce the Schiff moments by factors of two to four. Furthermore, the Schiff moments of $^{229}$Th and $^{229}$Pa are predicted to exceed that of $^{225}$Ra by more than an order of magnitude, largely due to the near-degenerate parity doublets predicted by MR-CDFT. This work establishes a microscopic framework for calculating Schiff moments in heavy octupole-deformed nuclei, providing essential nuclear-structure input for ongoing and future EDM searches.

\end{abstract}
 
\maketitle

\paragraph{\textbf{\textit{Introduction.}--}} The origin of the matter–antimatter asymmetry in the Universe remains one of the central unresolved problems in fundamental physics.  Its generation requires the violation of charge–parity (CP) symmetry~\cite{Sakharov:1967}. 
The amount of CP violation contained in the Standard Model, arising from the phase of the Cabibbo–Kobayashi–Maskawa (CKM) matrix~\cite{Hocker:2006},  is far too small to account for the observed baryon asymmetry, suggesting the presence of additional CP-violating sources beyond the Standard Model. 
Searches for permanent electric dipole moments (EDMs) of particles, atoms, and molecules provide one of the most sensitive probes of such new CP-violating interactions~\cite{Yamanaka:2015,Yamanaka:2017,Chupp:2017}. 
Existing experiments have already imposed stringent constraints on CP-violating phases in many extensions of the Standard Model~\cite{Graner:2016Hg199_EDM,ACME:2018,Abel:2020}, and the next generation of EDM experiments is expected to improve their sensitivity by several orders of magnitude~\cite{Alarcon:Snowmass2022}.

The most stringent atomic EDM limit is currently obtained for diamagnetic $^{199}$Hg, with
$|d_{\rm Hg}|<7.4\times10^{-30} e {\rm cm}$ at 95\% C.L. ~\cite{Graner:2016Hg199_EDM}, about four orders of magnitude above the
CKM-induced value~\cite{Yoshinaga:2018PTSM3}; competitive searches also
exist for \nuclide[129]{Xe}~\cite{Sachdeva:2019PRL}, \nuclide[171]{Yb}~\cite{Zheng:2022PRL}, and \nuclide[205]{Tl}~\cite{Regan:2002PRL}.  
In diamagnetic atoms, the dominant nuclear contribution arises from
the Schiff moment~\cite{Haxton:1983,Flambaum:1984,Flambaum:1986NPA}, induced by parity- and time-reversal-violating ($P,T$-odd) nuclear
interactions. Perturbatively, it is determined by products of the
$P,T$-odd and Schiff-operator matrix elements between the ground
state and opposite-parity excited states, weighted by inverse
excitation energies~\cite{Engel:2025}. The small parity-doublet
splittings generated by reflection-asymmetric shapes can therefore
strongly enhance Schiff moments, making $^{225}$Ra~\cite{Parker:2015Ra225_EDM},
$^{229}$Th, and $^{229}$Pa promising candidates for next-generation
EDM searches~\cite{Alarcon:Snowmass2022}.

Accurate calculations of nuclear Schiff moments in candidate atoms are essential for interpreting current searches and guiding future EDM experiments. This remains a major challenge for nuclear theory, as it requires a precise description of both ground state and opposite-parity excited states. At present, theoretical predictions of Schiff moments differ substantially among nuclear models~\cite{Spevak:1996PRM,Flambaum:1986NPA,Dmitriev:2003RPA2,Dmitriev:2004PRC,deJesus:2005SQRPA1,Dobaczewski:2005PRL,Ban:2010PRC,Teruya:2017PTSM1,Yoshinaga:2018PTSM3,Yoshinaga:2018PTSM4,Yanase:2020LSSM1}, leading to uncertainties exceeding an order of magnitude in the extracted CP-violating couplings. Reducing these nuclear-structure uncertainties is therefore crucial for translating EDM measurements into quantitative constraints on physics beyond the Standard Model~\cite{Chupp:2017,Engel:2013review1,Ginges:2003review,Engel:2025}.

Recently, ab initio methods have been extended to nuclear Schiff moments, but their applications remain limited to light nuclei~\cite{Ng:2026} or weakly deformed systems~\cite{Belley:2026}. Existing studies of Schiff moments in heavy deformed nuclei have relied either on particle-rotor models (PRM)~\cite{Auerbach:1996,Spevak:1996PRM,Flambaum:2014,Flambaum:2020,Sushkov:2024} or on nuclear energy-density-functional (EDF) theory~\cite{Dobaczewski:2005PRL,Dobaczewski:2018PRL}. Both approaches commonly assume that the ground and opposite-parity excited states share the same rigid intrinsic shape. Mean-field EDF calculations capture important deformation effects through symmetry-breaking intrinsic states, but they may miss dynamical correlations associated with symmetry restoration and shape fluctuations. These correlations can strongly affect both ground-state properties and the low-lying opposite-parity states that dominate Schiff moments. A natural framework for incorporating such effects is provided by beyond-mean-field methods based on the generator coordinate method (GCM)~\cite{Ring:1980}, which mix configurations with different intrinsic shapes and restore broken symmetries.

In this Letter, we extend multireference covariant density functional theory (MR-CDFT) to nuclear Schiff moments in the heavy octupole-deformed nuclei \nuclide[225]{Ra}, \nuclide[229]{Th}, and \nuclide[229]{Pa}. The matrix elements of the pion-exchange $P,T$-odd two-body nucleon--nucleon ($NN$) interaction and electromagnetic operators are evaluated in a fully relativistic formulation, with no adjustable parameters beyond those of the underlying universal EDF. Beyond-mean-field correlations arising from symmetry restoration and shape mixing are incorporated through the GCM, with projection onto states of good parity, particle number, and angular momentum. We show that dynamical shape fluctuations substantially modify the predicted Schiff moments of octupole-deformed nuclei, reducing them by factors of two to four. This study establishes a microscopic and universal framework for calculating Schiff moments in heavy octupole-deformed nuclei, providing essential nuclear-structure input for the interpretation of future EDM experiments.

\paragraph{\textbf{\textit{Formalism.--}}} The Schiff moment induced by the $P,T$-odd two-body $NN$ interaction, $\hat{V}_{\rm PT}$~\cite{Haxton:1983,Maekawa:2011}, originates from the following pion–nucleon Lagrangian density,
\begin{eqnarray} \mathcal{L}^{\rm PVTV}_{\pi NN} &=&\bar g^{(0)}_{\pi NN}\bar N \vec{\tau}\cdot\vec{\pi}N +\bar g^{(1)}_{\pi NN}\bar N N \pi_z \nonumber\\ &&+\bar g^{(2)}_{\pi NN}\bar N (3\tau_z\pi_z-\vec{\tau}\cdot\vec{\pi}) N, \end{eqnarray}
where $N$ denotes the nucleon fields, $\vec{\tau}$ are the isovector Pauli matrices, and $\vec{\pi}$ represents the pion fields. The superscripts 
0, 1, and 2 denote the isoscalar, isovector, and isotensor terms, respectively. The three LECs $\bar g^{(\alpha)}_{\pi NN}$ are unknown new physics parameters. The resulting Schiff moment can then be evaluated within second-order perturbation theory~\cite{Flambaum:1984,Engel:2013review1}
\begin{eqnarray} 
\label{eq:Schiff_moment_def}
S  &=& \sum_{k\ne 0} \cfrac{\langle \Psi_0|\hat S_z|\Psi_k\rangle\langle \Psi_k|{{\hat V_{\rm PT}}}|\Psi_0\rangle}{E_0-E_k}+c.c. \nonumber\\
& \equiv &  g_{\pi NN} \sum_{\alpha=0}^2 \bar g^{(\alpha)}_{\pi NN}  a_\alpha  \,.
\end{eqnarray} 
Here the abbreviation $c.c.$\ denotes the complex conjugate, and the states $\ket{\Psi_0}$ and $\ket{\Psi_k}$ are the ground state and the $k$-th excited state of the unperturbed strong Hamiltonian, having the same total angular momentum but opposite parity, with  $E_0$ and $E_k$ the corresponding energies.  
All nuclear-structure information is contained in the ``structure factors'' $a_\alpha$. 
The Schiff operator $\hat S_z$  is~\cite{Flambaum:1984}
\begin{eqnarray}
\hat{S}_z= \cfrac{e}{10}\sum_p\left( 
\hat{r}_p^2-\cfrac{5}{3}~r_{\rm{ch}}^2
\right)\hat{z}_p \,,
\end{eqnarray}
where  the sum runs over protons, $r^2_{\rm{ch}} $ is the mean-square charge radius, and we have omitted a smaller quadrupole correction and contributions from the nucleon EDMs.
The Lagrangian density for the strong unperturbed $P,T$-conserving $\pi NN$ coupling is 
\begin{eqnarray}
  \mathcal{L}^{PCTC}_{\pi NN} &=&i g_{\pi NN}\bar N \gamma_5 \vec{\tau}\cdot\vec{\pi}N,
\end{eqnarray}
where the standard $\pi NN$ coupling constant is determined by the Goldberger–Treiman theorem $g_{\pi NN}=m_Ng_A/f_\pi\simeq 12.9$~\cite{Goldberger:1958}. In Eq.~(\ref{eq:Schiff_moment_def}), the matrix elements of the two-body $P,T$-odd nuclear interaction, $\bra{\Psi_k} \hat V_{\rm PT} \ket{\Psi_0}$, are evaluated explicitly within a fully relativistic framework, following an approach analogous to that used in neutrinoless double-$\beta$ ($0\nu\beta\beta$) decay calculations, as detailed in Ref.~\cite{Song:2014}. 

The many-body wave functions of nuclear states are constructed as the superposition of symmetry-conserved basis functions~\cite{Zhou:2024PRC,Zhou:2025_Long}
\begin{equation}
\label{eq:gcmwf}
\ket{ \Psi^{J\pi}_k}
=\sum_{c} f^{J k\pi}_{c} \ket{NZ J\pi; c}.
\end{equation}
Here, $k$ distinguishes states with the same angular momentum $J$, and the symbol $c$ is a collective label for the indices $(\kappa,\mathbf{q})$ that label the blocked-quasiparticle level and deformation parameters, respectively. The basis function with correct quantum numbers ($NZJ\pi$) is given by
\begin{equation}
\label{eq:basis}
\ket{NZ J\pi; c} 
=  \hat P^J_{MK} \hat P^N\hat P^Z\hat P^\pi\ket{\Phi^{\rm (odd)}_\kappa(\mathbf{q})},
\end{equation}
where "odd" stands for ``odd number of nucleons'' and the corresponding states are mean-field configurations of odd-mass nuclei; $\hat P^{J}_{MK}$, $\hat{P}^{N, Z}$, and $\hat P^\pi$ are projection operators that select components with the angular momentum $J$, $z$-projection $M$, neutron number $N$, proton number $Z$, and parity $\pi=\pm$~\cite{Ring:1980}.

The weight function $f_{c}^{Jk\pi}$ in Eq.~(\ref{eq:gcmwf}) is determined variationally through the Hill--Wheeler--Griffin equation~\cite{Hill:1953,Griffin:1957,Ring:1980}. Solving this equation yields the energies and collective wave functions of the ground and excited states. These wave functions allow for the calculation of a variety of observables, including electric multipole transition strengths, magnetic dipole transition strengths, and magnetic dipole moments, which provide important benchmarks for validating the MR-CDFT description of Schiff moments. Since the full single-particle basis is employed, no effective charges or renormalized neutron and proton $g$ factors are introduced.

\paragraph{\textbf{\textit{Numerical details.--}}}
The mean-field wave functions in Eq.(\ref{eq:basis}) are chosen as one-quasiparticle  configurations,
$\ket{\Phi^{\rm (odd)}_\kappa(\mathbf{q})}=\alpha^\dagger_\kappa \ket{\Phi_{(\kappa)}(\mathbf{q})}$,
where $\kappa$ labels the blocked orbital. The quasiparticle vacuum states $\ket{\Phi_{(\kappa)}(\mathbf{q})}$ are generated from self-consistent CDFT  calculations~\cite{Meng:2016Book,Zhao:2010PRC,Yao:2015Ra224} with constraints on the odd nucleon number as well as on the mass quadrupole and octupole moments,
\beq 
Q_{\lambda 0}=\bra{\Phi_{(\kappa)}(\mathbf{q})} r^\lambda Y_{\lambda 0}\ket{\Phi_{(\kappa)}(\mathbf{q})},
\eeq 
where $\mathbf{q}=(\beta_2,\beta_3)$ denotes the quadrupole and octupole deformation parameters~\cite{Zhou:2024PRC}, defined by
\beq
\beta_\lambda \equiv \frac{4\pi}{3AR^\lambda}Q_{\lambda 0},
\eeq
with $R=1.2A^{1/3}$ fm and $A$ the mass number.
To remove spurious center-of-mass motion, we fix the center of mass at the origin by imposing the constraint $Q_{10}=0$ on the mean-field configurations~\cite{Yao:2015Ra224,Yao:2016Pb208}. This constraint essentially eliminates the matrix elements of the center-of-mass operator between the ground state and excited states.

Axial symmetry is imposed, so the mean-field configurations are characterized by the projection $K$ of angular momentum along the intrinsic $z$ axis. The Dirac equation is solved in a
harmonic-oscillator (HO) basis containing the first 13 major shells ($N_{\rm sh}=12$), with $\hbar\omega_0=41A^{-1/3}$ MeV, which
provides reasonable convergence for low-lying excitation energies
and spectroscopic properties. We use the PC-PK1 EDF~\cite{Zhao:2010PRC}
and treat pairing in the BCS approximation with a $\delta$ interaction and smooth energy cutoff~\cite{Yao:2010}. Angular-momentum
and particle-number projections use 16 rotational-angle and 7 gauge-angle mesh points. Although symmetry restoration with modern
EDFs can generate spurious divergences~\cite{Anguiano:2001,Tajima:1992,Dobaczewski:2007} and finite-step
artifacts~\cite{Bender:2009,Duguet:2009}, these effects are expected
to be minor for the present mesh parameters~\cite{Zhou:2024PRC}.
The framework has previously been applied successfully to low-lying
quadrupole- and octupole-correlated states~\cite{Yao:2016Pb208,Yao:2015Ra224};
further details are given in Refs.~\cite{Yao:2010,Zhou:2024PRC,Zhou:2025,Zhou:2025_Long}. 

\paragraph{\textbf{\textit{Results and discussion.--}}} Before presenting the Schiff-moment results, we briefly summarize the key nuclear-structure properties predicted by MR-CDFT. The quadrupole--octupole deformations $(\beta^{\rm min}_2,\beta^{\rm min}_3)$ at the mean-field energy minima are $(0.18,0.11)$ for \nuclide[225]{Ra}, $(0.22,0.14)$ for \nuclide[229]{Th}, and $(0.22,0.15)$ for \nuclide[229]{Pa}. After symmetry restoration, the ground-state energy minima shift slightly to $(0.18,0.12)$ for the $J^\pi=1/2^+$ ground state of \nuclide[225]{Ra}, and to $(0.22,0.16)$ and $(0.25,0.16)$ for the $J^\pi=5/2^+$ ground states of \nuclide[229]{Th} and \nuclide[229]{Pa}, respectively~\cite{Zhou:2025_Long}. These deformation parameters are significantly larger than the values $(0.143,0.099)$ for \nuclide[225]{Ra} and $(0.176,0.082)$ for \nuclide[229]{Pa} adopted in previous particle-rotor-model (PRM) studies~\cite{Cwiok:1991,Spevak:1996PRM}.

The calculated magnetic dipole moments of the ground states of \nuclide[225]{Ra} and \nuclide[229]{Th} are $-0.783 \mu_N$ and $+0.32 \mu_N$, respectively, in good agreement with the experimental values $-0.734 \mu_N$ and $+0.360(7)  \mu_N$~\cite{NNDC}. For the isomeric $3/2^+$ state in \nuclide[229]{Th}, MR-CDFT predicts a magnetic dipole moment of $-0.48 \mu_N$, and spectroscopic quadrupole moment $Q_s = 171 e$ fm$^2$, compared with the data $\mu=-0.37(6)\mu_N$ and $Q_s=174(6) e$ fm$^2$ \cite{NNDC}, respectively, whose reproduction has proven challenging for most nuclear models \cite{Minkov:2024}. The calculated $B(E2; 5/2^+_1 \rightarrow 1/2^+_1)$ strength in \nuclide[225]{Ra} is $101$ Weisskopf units (W.u.), in excellent agreement with the experimental value of $102(6)$ W.u.~\cite{NNDC}. Additional details can be found in Ref.~\cite{Zhou:2025_Long}, where we show that the electromagnetic properties of the low-lying states in \nuclide[225]{Ra}, \nuclide[229]{Th}, and \nuclide[229]{Pa} are generally reproduced at the $\sim10\%$ level. These results provide important benchmarks for the reliability of the present framework in describing the collective structure of heavy octupole-deformed nuclei.

    \begin{figure}[bt]
 \centering 
\includegraphics[width=\columnwidth]{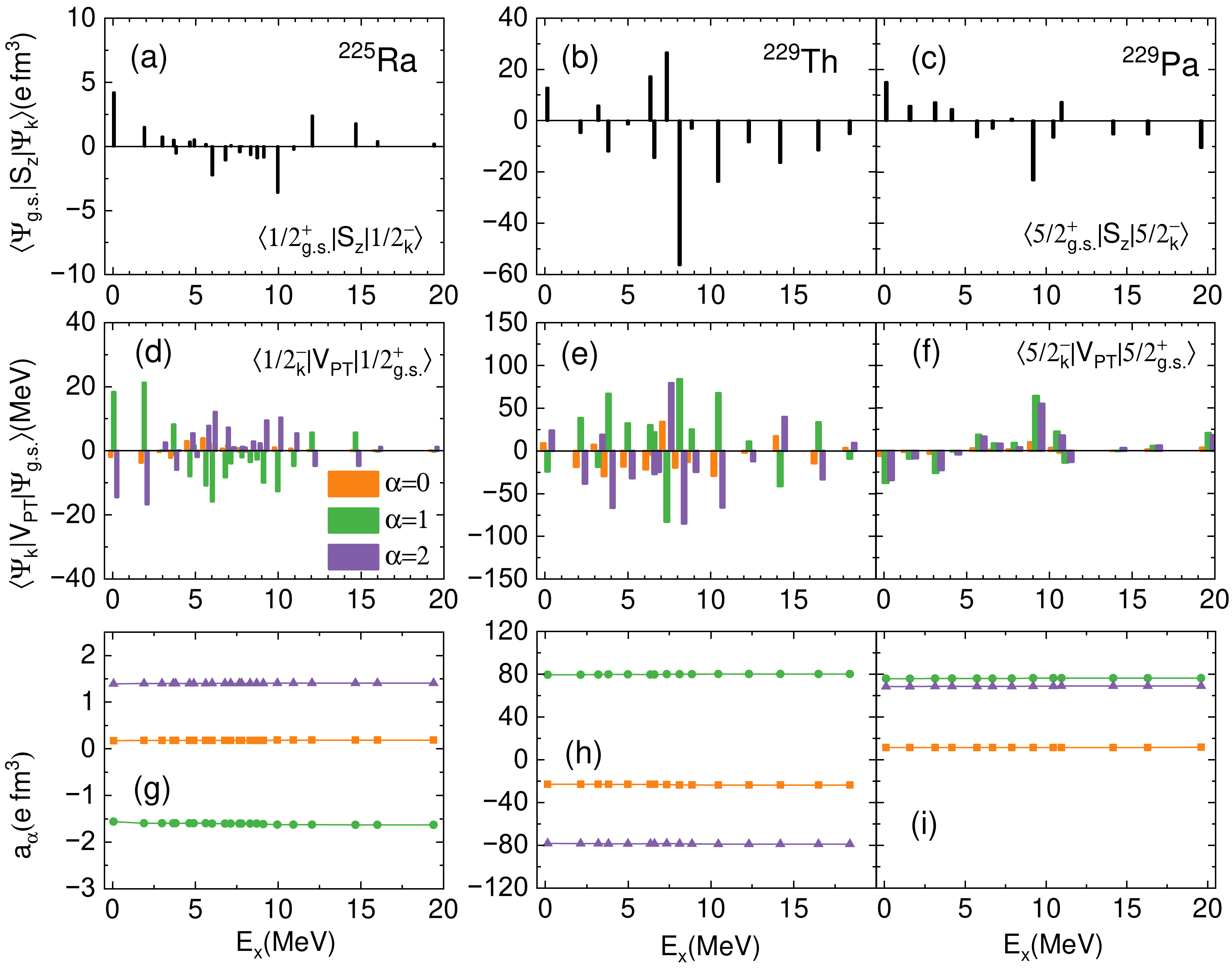}
\caption{(Color online) Matrix elements of the Schiff operator $\hat S_z$ and the $P,T$-violating $NN$ interaction $\hat V_{\rm PT}$ between the ground state and opposite-parity excited states with the same angular momentum, together with the accumulated structure factors $a_\alpha$ from the MR-CDFT calculations. Results are shown as functions of the excitation energy $E_x$ for (a,d,g) \nuclide[225]{Ra}, (b,e,h) \nuclide[229]{Th}, and (c,f,i) \nuclide[229]{Pa}. Contributions from the three components of $\hat V_{\rm PT}$ with $\alpha=0,1$, and $2$ are shown in orange, green, and purple, respectively, with slight horizontal offsets for clarity.}
 \label{fig:Sz_Vpt_ME}
 \end{figure}

Figure~\ref{fig:Sz_Vpt_ME} displays the calculated ground-to-excited-state matrix elements of the Schiff operator $\hat S_z$ and the $P,T$-odd $NN$ interaction $\hat V_{\rm PT}$, together with their contributions to the structure factors $a_\alpha$ in the three isotopes. Although sizable matrix elements of both $\hat S_z$ and $\hat V_{\rm PT}$ appear in the low- and high-energy regions, only the first excited state gives a significant contribution in Eq.~(\ref{eq:Schiff_moment_def}). Contributions from higher-lying states are strongly suppressed by the energy denominator. 
Quantitatively, the first excited state accounts for more than 94\%, 96\%, and 99\% of the total $a_\alpha$ values in \nuclide[225]{Ra}, \nuclide[229]{Th}, and \nuclide[229]{Pa}, respectively. This confirms that the Schiff moment of heavy octupole-deformed nuclei is governed primarily by the ground state and the first opposite-parity excited state with the same angular momentum. 

Our MR-CDFT framework enables a direct test of the validity of the PRM formula for nuclear Schiff moments in octupole-deformed nuclei, given by~\cite{Spevak:1996PRM,Flambaum:2020}
\begin{eqnarray}
\label{eq:Schiff_moment_PRM}
S_{\rm PRM}
&\simeq& 2 \frac{S_0}{\Delta E_{\pm}} \sum_{\alpha=0,1,2} V^{(\alpha)}_{\rm PT},
\end{eqnarray}
where $\Delta E{\pm}=|E(J^-)-E(J^+)|$ is the energy splitting between the parity-doublet states, in units of keV. The relevant matrix elements of the Schiff operator and the $\hat V_{\rm PT}$ are approximated as~\cite{Auerbach:1996,Spevak:1996PRM,Flambaum:2014,Flambaum:2020,Sushkov:2024}
\bsub\beqn
\label{eq:S0}
S_0
&\equiv& \langle J^+ | \hat S_{z}|J^-\rangle
\simeq \frac{J}{J+1}\frac{3}{20 \pi \sqrt{35}} e Z R^3 \beta_2 \beta_3,\\
\label{eq:VPT}
\sum_{\alpha}V^{(\alpha)}_{\rm PT} &\equiv& \langle J^- | \hat V_{\rm PT}|J^+\rangle
\simeq 10^{-3}\beta_3 A^{-1/3}\eta(\bar g^{(\alpha)}_{\pi NN}),
\eeqn
\esub
with $R=1.2 A^{1/3}$~fm. In the derivation of the above expression, several approximations are used for the wave functions and for the $P,T$-odd $NN$ interaction $\hat V_{\rm PT}$~\cite{Auerbach:1996,Spevak:1996PRM}, with the  dimensionless strength parameter  $\eta$ approximated as $\eta_n\simeq 5\times 10^6 g_{\pi NN}\left(
-0.2\bar{g}_{\pi NN}^{(0)}
+ \bar{g}_{\pi NN}^{(1)}
+ 0.4\bar{g}_{\pi NN}^{(2)}
\right)$ \cite{Flambaum:2014}.

 \begin{figure}[bt]
 \centering 
\includegraphics[width=8.6cm]{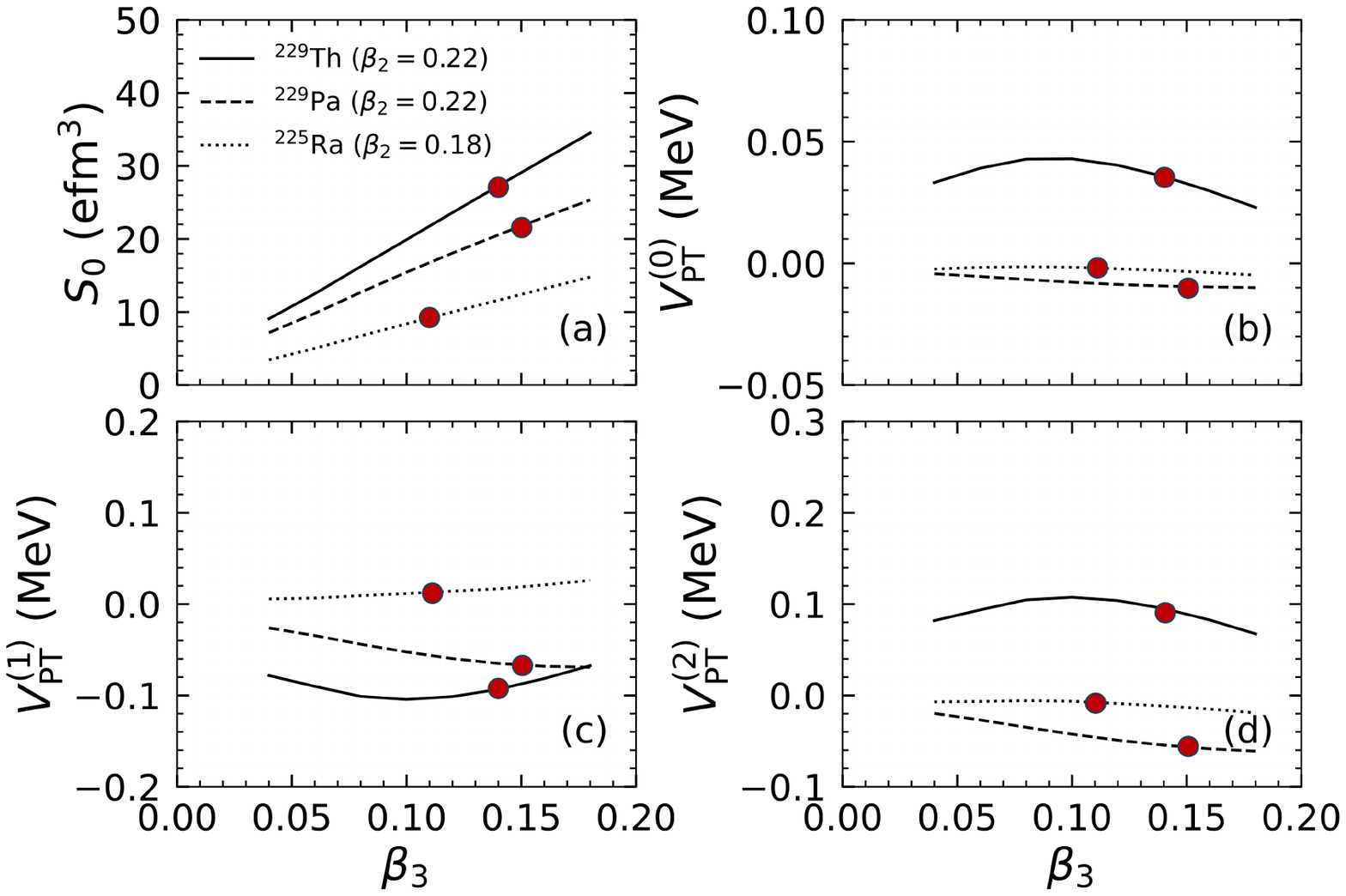} 
\caption{Matrix elements of (a) $\bra{J^+}\hat S_z\ket{J^-}$ and (b--d) $\bra{J^-}\hat V^{(\alpha)}_{\rm PT}\ket{J^+}$ with $\alpha=0,1,2$, respectively, between the ground state $(J^+)$ and the first opposite-parity excited state $(J^-)$ in \nuclide[225]{Ra}, \nuclide[229]{Th}, and \nuclide[229]{Pa}. The results are obtained from the MR-CDFT calculations with the single-state (SS) and single-configuration (SC) approximation with $\eMax=12$, where the quadrupole deformation $\beta_2$ of the configuration is fixed to be that of the energy-minimum mean-field configuration (indicated by red dots), while varying the octupole deformation $\beta_3$, followed by projection onto good quantum numbers $(N,Z,J^\pi)$.   
}
\label{fig:S0_VPT_beta3}
 \end{figure}

Figure~\ref{fig:S0_VPT_beta3}(a) shows $S_0$ as functions of the octupole deformation $\beta_3$, with  $\beta_2$ fixed at their mean-field energy-minimum values. In all three nuclei, $S_0$ is
positive and increases linearly with $\beta_3$, validating  Eq.~(\ref{eq:S0}). The slopes are larger in $^{229}$Th and $^{229}$Pa because of their larger $\beta_2$, $Z$, and $A$. By contrast, the three $V_{PT}^{(\alpha)}$ matrix elements have a more complicated dependence on $\beta_3$, varying in magnitude and sign according to the isospin channel. In $^{229}$Th, a kink near $\beta_3=0.10$
reflects a change in the underlying configuration. 
 The $P,T$-odd two-body matrix elements are therefore more sensitive than $S_0$ to the intrinsic configuration, challenging rigid-shape approximations.  The energy-minimum configurations indicated in Fig.~\ref{fig:S0_VPT_beta3} show the
largest matrix elements in $^{229}$Th among the three isotopes.

\begin{table}[tb]
\centering
\tabcolsep=5pt
\caption{Structure factors $a_\alpha$ ($e\mathrm{fm}^3$) for \nuclide[225]{Ra}, \nuclide[229]{Th}, and \nuclide[229]{Pa}, obtained from full MR-CDFT calculations extrapolated to $\eMax \to \infty$. The results are compared with those from the single-state (SS) approximation, in which only the first opposite-parity intermediate state is included and its experimental excitation energy $\Delta E_{\pm}$ is used. Also shown for comparison are SS results based on the single-configuration  (SS+SC) approximation, as well as those from the pure PRM, both using the experimental value of $\Delta E_{\pm}$. In the SS+SC calculation, the mean-field energy-minimum configuration is used. In the PRM calculation, the deformation parameters $\beta_2$ and $\beta_3$ are taken, for simplicity, from the corresponding mean-field energy-minimum configuration.} 
\begin{tabular}{llrrrr}
\hline
Method & Nucleus  & $a_0$ & $a_1$ & $a_2$ & $\Delta E_{\pm}$ (keV)\\
\hline
\multirow{3}{*}{\makecell{MR-CDFT\\(Full)}} & $^{225}\mathrm{Ra} $  & $+0.11$ & $-1.3$ & $+0.92$ & $78$\\
  & $^{229}\mathrm{Th}$  & $-24$   & $+80$  & $-79$ &  $15$\\
& $^{229}\mathrm{Pa}$  & $+12$   & $+76$  & $+69$&   $16$ \\
\hline
\multirow{3}{*}{\makecell{MR-CDFT\\(SS)}} & $^{225}\mathrm{Ra}$  & $+0.15$ & $-1.9$ & $+1.3$ & $55$\\
 & $^{229}\mathrm{Th}$& $-3.3$   & $+8.8$ & $-7.7$  & $146$ \\
& $^{229}\mathrm{Pa}$ & $+1.8$  & $+12$  & $+10$ & $99$  \\
\hline  
 &   &  &   &   & $(\beta^{\rm min}_2, \beta^{\rm min}_3)$ \\
\hline
\multirow{3}{*}{\makecell{MR-CDFT\\(SS+SC)}} & $^{225}\mathrm{Ra}$  & $+0.66$ & $-4.4$ & $+2.6$ & $(0.18, 0.11)$\\
 & $^{229}\mathrm{Th}$& $-13$   & $+35$ & $-36$  & $(0.22, 0.14)$ \\
& $^{229}\mathrm{Pa}$ & $+4.2$  & $+29$  & $+25$ & $(0.22, 0.15)$  \\
\hline 
& $^{225}\mathrm{Ra}$  & $-4.3$ & $+21.4$ & $+8.6$ & $(0.18, 0.11)$\\
PRM & $^{229}\mathrm{Th}$& $-7.2$   & $+36$ & $+14$  & $(0.22, 0.14)$ \\
& $^{229}\mathrm{Pa}$ & $+12$  & $-62$  & $-25$ & $(0.22, 0.15)$   \\
\hline 

\end{tabular} 
    \label{tab:schiff_moments}
\end{table}

The structure factors obtained from the full MR-CDFT calculations, including symmetry restoration and shape-fluctuation effects, for \nuclide[225]{Ra}, \nuclide[229]{Th}, and \nuclide[229]{Pa} are listed in Table~\ref{tab:schiff_moments}. Since both the Schiff-moment operator $\hat S$ and the $P,T$-violating interaction $\hat V_{\rm PT}$ are long-range operators, their matrix elements converge slowly with respect to the HO basis cutoff $\eMax$. To quantify and reduce the associated truncation uncertainty, we perform calculations with $\eMax=8$, 10, and 14, and extrapolate the resulting structure factors $a_\alpha$ to the limit $\eMax\rightarrow\infty$.

For comparison, we also present results obtained at different levels of approximation, namely MR-CDFT (SS) and MR-CDFT (SS+SC). In these calculations, only the lowest opposite-parity excited state is retained in Eq.(\ref{eq:Schiff_moment_def}), and its excitation energy is taken from experiment~\cite{NNDC}. As shown in Table~\ref{tab:schiff_moments}, the structure factors $a_\alpha$ obtained from the full MR-CDFT calculations for \nuclide[225]{Ra} are comparable to those from the MR-CDFT (SS) approximation. In contrast, the corresponding values for \nuclide[229]{Th} and \nuclide[229]{Pa} are approximately one order of magnitude larger than the MR-CDFT (SS) results. In light of the results presented in Fig.~\ref{fig:Sz_Vpt_ME}, which demonstrate that the full MR-CDFT structure factors are overwhelmingly dominated by the contribution of the lowest opposite-parity state, this discrepancy can be attributed primarily to the energy denominator $\Delta E_\pm$. 

We note that MR-CDFT predicts a nearly degenerate negative-parity band as the parity partner of the ground-state band in both nuclei~\cite{Zhou:2025_Long}, although these states have not yet been observed experimentally. The predicted excitation energies of the bandhead ($5/2^-$) states are approximately 15 keV and 16 keV in \nuclide[229]{Th} and \nuclide[229]{Pa}, respectively. These values are substantially lower than those of the experimentally observed first opposite-parity excited states, which are located at approximately 146 keV and 99 keV, respectively~\cite{NNDC}. As discussed in Refs.~\cite{Levon:1994,Zhou:2025_Long} and adopted in the NNDC assignment~\cite{NNDC}, the experimentally observed first $5/2^-$ state in \nuclide[229]{Pa} is dominated by the Nilsson configuration \cite{Ring:1980} of $\pi 1/2[530]$, whereas the ground state is dominated by $\pi 5/2[642]$. Consequently, the observed $5/2^-$ state does not belong to the parity-doublet structure built on the ground-state band and therefore cannot be regarded as its parity partner. Given the present theoretical uncertainties in MR-CDFT predictions of excitation energies, the actual energy splitting of the ground-state parity doublet may be even smaller than predicted here. Indeed, an early study of $\gamma$-ray and conversion-electron transitions following the electron-capture decay of \nuclide[229]{U} reported an excitation energy of $0.22\pm0.05$ keV in \nuclide[229]{Pa}~\cite{Ahmad:1982}, although no corresponding state was identified in the later $^{230}{\rm Th}(p,2n)^{229}{\rm Pa}$ and $^{231}{\rm Pa}(p,t)^{229}{\rm Pa}$ experiments~\cite{Grafen:1991}. Such an exceptionally small parity-doublet splitting could enhance the Schiff moment by several additional orders of magnitude. Although this intriguing possibility remains to be established experimentally, a renewed search for these predicted near-degenerate states would be of considerable interest. Their observation is expected to be extremely challenging, but may become feasible with advanced laser-spectroscopy techniques~\cite{Beck:2007,Seiferle:2019,Yamaguchi:2019}.

The difference between MR-CDFT (SS) and MR-CDFT (SS+SC) illustrates the impact of shape fluctuation effect. Both use the same experimental energy denominator, but MR-CDFT (SS+SC) evaluates the matrix elements of $\hat S$ and $\hat V_{\rm PT}$ using only the mean-field energy-minimum configuration. Shape mixing is found to quench the structure factors $a_\alpha$ by a factor of 2--4. This quenching originates from the mismatch between the squared collective wave functions of the parity-doublet states \cite{Zhou:2025_Long}. The dominant configurations of the opposite-parity states differ in distinct ways across the three nuclei: in \nuclide[225]{Ra}, the excited state has larger $(\beta_2,\beta_3)$ values than the ground state; in \nuclide[229]{Th}, it has a larger $\beta_3$ but smaller $\beta_2$; and in \nuclide[229]{Pa}, $\beta_3$ decreases slightly while $\beta_2$ remains nearly unchanged. These results indicate that the rigid-shape approximation is insufficient for a reliable description of Schiff moments even in heavy octupole-deformed nuclei. Moreover, we find that, within MR-CDFT (SS+SC), pure symmetry-restoration effects on nuclear Schiff moments are minor, consistent with previous findings~\cite{Dobaczewski:2018PRL}. To estimate the uncertainty associated with the choice of EDF, we repeated the calculations using the widely employed PC-F1 functional~\cite{Burvenich:2002PRC}. The resulting structure factors differ by approximately 30\%--50\%, providing an indication of the systematic uncertainty related to the EDF. 

Finally, we also compare MR-CDFT (SS+SC) with results obtained from the PRM formula in Eq.~(\ref{eq:Schiff_moment_PRM}), using deformation parameters of the corresponding mean-field minima. This comparison tests the approximations underlying the PRM formula. The structure factors obtained with the two approaches can differ by up to a factor of six and may even have opposite signs. This is because the relative signs among different $a_\alpha$ are fixed in the PRM \cite{Flambaum:2014}, whereas they are nucleus-dependent in MR-CDFT (SS+SC). These relative signs are important for global analyses of constraints on the  LECs from EDM measurements~\cite{Chupp:2015}.

\paragraph{\textbf{\textit{Conclusions.}--}} We have extended MR-CDFT to nuclear Schiff moments and performed fully microscopic beyond-mean-field study of the heavy octupole-deformed nuclei \nuclide[225]{Ra}, \nuclide[229]{Th}, and \nuclide[229]{Pa}, which are among the most promising candidates for EDM searches. Based on a universal relativistic energy density functional (EDF), the framework incorporates shape fluctuations and symmetry restoration within the generator coordinate method, with projection onto states of good parity, particle number, and angular momentum. Without introducing additional free parameters beyond those in the EDF, it captures the essential features of the low-energy structure of these octupole-deformed nuclei and reproduces electromagnetic observables reasonably well. Our calculations show that symmetry-restoration effects on nuclear Schiff moments are minor, whereas shape fluctuations reduce the Schiff moments by factors of two to four. After incorporating these effects in the full MR-CDFT framework, the Schiff moments of \nuclide[229]{Th} and \nuclide[229]{Pa} are predicted to be more than an order of magnitude larger than that of \nuclide[225]{Ra}. This enhancement is driven primarily by their much smaller energy denominators associated with low-lying opposite-parity states, which remain to be confirmed experimentally. These results provide essential nuclear-structure input for ongoing and future searches for atomic and molecular EDMs.

It is worth noting that, within nonrelativistic chiral effective field theory, a CP-violating contact interaction between nucleons appears at leading order~\cite{deVries:2020l}. Whether an analogous contribution enters at the same order in a fully relativistic formulation remains an open question. A complete quantification of the systematic and parametric uncertainties in the predicted Schiff moments also remains to be carried out. While systematic uncertainties are challenging to assess within the present framework, parametric uncertainties can be quantified using the recently developed subspace-projected CDFT approach~\cite{Zhang:2024_Letter}. Nevertheless, these refinements are not expected to alter the main conclusion that shape mixing significantly quenches Schiff moments in heavy octupole-deformed nuclei. The present framework can also be applied straightforward to other heavy octupole-deformed systems, such as \nuclide[223]{Rn}, \nuclide[223]{Ra}, \nuclide[221,223]{Fr}, \nuclide[227]{Th} and \nuclide[225,227]{Ac} \cite{Athanasakis-Kaklamanakis:2025},  enabling systematic predictions of their Schiff moments within a unified approach. Such global calculations will be important for extracting consistent constraints on new-physics parameters from EDM measurements in different atomic and molecular systems.

\paragraph{Acknowledgments.} 
We thank H. Hergert, X. Fan, Z. C. Gao, Z.T. Lu, J. T. Singh, A. Vutha, C.C. Wang,  T. Xia,  and X. F. Yang for helpful discussions, N. Yamanaka and K. Yanase for careful reading of the manuscript, and the hospitality at APCTP where part of this work was done. This work is supported in part by the National Natural Science Foundation of China (Grant Nos. 12405143, 12375119 and 12141501).  JE acknowledges support from the US Department of Energy under award DE-FG02-97ER41019 and its Topical Collaboration in Nuclear Theory for New Physics, award No. DE-SC0023663.


\begin{thebibliography}{69}%
\makeatletter
\providecommand \@ifxundefined [1]{%
 \@ifx{#1\undefined}
}%
\providecommand \@ifnum [1]{%
 \ifnum #1\expandafter \@firstoftwo
 \else \expandafter \@secondoftwo
 \fi
}%
\providecommand \@ifx [1]{%
 \ifx #1\expandafter \@firstoftwo
 \else \expandafter \@secondoftwo
 \fi
}%
\providecommand \natexlab [1]{#1}%
\providecommand \enquote  [1]{``#1''}%
\providecommand \bibnamefont  [1]{#1}%
\providecommand \bibfnamefont [1]{#1}%
\providecommand \citenamefont [1]{#1}%
\providecommand \href@noop [0]{\@secondoftwo}%
\providecommand \href [0]{\begingroup \@sanitize@url \@href}%
\providecommand \@href[1]{\@@startlink{#1}\@@href}%
\providecommand \@@href[1]{\endgroup#1\@@endlink}%
\providecommand \@sanitize@url [0]{\catcode `\\12\catcode `\$12\catcode
  `\&12\catcode `\#12\catcode `\^12\catcode `\_12\catcode `\%12\relax}%
\providecommand \@@startlink[1]{}%
\providecommand \@@endlink[0]{}%
\providecommand \url  [0]{\begingroup\@sanitize@url \@url }%
\providecommand \@url [1]{\endgroup\@href {#1}{\urlprefix }}%
\providecommand \urlprefix  [0]{URL }%
\providecommand \Eprint [0]{\href }%
\providecommand \doibase [0]{http://dx.doi.org/}%
\providecommand \selectlanguage [0]{\@gobble}%
\providecommand \bibinfo  [0]{\@secondoftwo}%
\providecommand \bibfield  [0]{\@secondoftwo}%
\providecommand \translation [1]{[#1]}%
\providecommand \BibitemOpen [0]{}%
\providecommand \bibitemStop [0]{}%
\providecommand \bibitemNoStop [0]{.\EOS\space}%
\providecommand \EOS [0]{\spacefactor3000\relax}%
\providecommand \BibitemShut  [1]{\csname bibitem#1\endcsname}%
\let\auto@bib@innerbib\@empty
\bibitem [{\citenamefont {Sakharov}(1967)}]{Sakharov:1967}%
  \BibitemOpen
  \bibfield  {author} {\bibinfo {author} {\bibfnamefont {A.~D.}\ \bibnamefont
  {Sakharov}},\ }\href {\doibase 10.1070/PU1991v034n05ABEH002497} {\bibfield
  {journal} {\bibinfo  {journal} {Pisma Zh. Eksp. Teor. Fiz.}\ }\textbf
  {\bibinfo {volume} {5}},\ \bibinfo {pages} {32} (\bibinfo {year}
  {1967})}\BibitemShut {NoStop}%
\bibitem [{\citenamefont {Hocker}\ and\ \citenamefont
  {Ligeti}(2006)}]{Hocker:2006}%
  \BibitemOpen
  \bibfield  {author} {\bibinfo {author} {\bibfnamefont {A.}~\bibnamefont
  {Hocker}}\ and\ \bibinfo {author} {\bibfnamefont {Z.}~\bibnamefont
  {Ligeti}},\ }\href {\doibase 10.1146/annurev.nucl.56.080805.140456}
  {\bibfield  {journal} {\bibinfo  {journal} {Ann. Rev. Nucl. Part. Sci.}\
  }\textbf {\bibinfo {volume} {56}},\ \bibinfo {pages} {501} (\bibinfo {year}
  {2006})},\ \Eprint {http://arxiv.org/abs/hep-ph/0605217}
  {arXiv:hep-ph/0605217} \BibitemShut {NoStop}%
\bibitem [{\citenamefont {Yamanaka}\ and\ \citenamefont
  {Hiyama}(2016)}]{Yamanaka:2015}%
  \BibitemOpen
  \bibfield  {author} {\bibinfo {author} {\bibfnamefont {N.}~\bibnamefont
  {Yamanaka}}\ and\ \bibinfo {author} {\bibfnamefont {E.}~\bibnamefont
  {Hiyama}},\ }\href {\doibase 10.1007/JHEP02(2016)067} {\bibfield  {journal}
  {\bibinfo  {journal} {JHEP}\ }\textbf {\bibinfo {volume} {02}},\ \bibinfo
  {pages} {067} (\bibinfo {year} {2016})},\ \Eprint
  {http://arxiv.org/abs/1512.03013} {arXiv:1512.03013 [hep-ph]} \BibitemShut
  {NoStop}%
\bibitem [{\citenamefont {Yamanaka}\ \emph {et~al.}(2017)\citenamefont
  {Yamanaka}, \citenamefont {Sahoo}, \citenamefont {Yoshinaga}, \citenamefont
  {Sato}, \citenamefont {Asahi},\ and\ \citenamefont {Das}}]{Yamanaka:2017}%
  \BibitemOpen
  \bibfield  {author} {\bibinfo {author} {\bibfnamefont {N.}~\bibnamefont
  {Yamanaka}}, \bibinfo {author} {\bibfnamefont {B.~K.}\ \bibnamefont {Sahoo}},
  \bibinfo {author} {\bibfnamefont {N.}~\bibnamefont {Yoshinaga}}, \bibinfo
  {author} {\bibfnamefont {T.}~\bibnamefont {Sato}}, \bibinfo {author}
  {\bibfnamefont {K.}~\bibnamefont {Asahi}}, \ and\ \bibinfo {author}
  {\bibfnamefont {B.~P.}\ \bibnamefont {Das}},\ }\href {\doibase
  10.1140/epja/i2017-12237-2} {\bibfield  {journal} {\bibinfo  {journal} {Eur.
  Phys. J. A}\ }\textbf {\bibinfo {volume} {53}},\ \bibinfo {pages} {54}
  (\bibinfo {year} {2017})},\ \Eprint {http://arxiv.org/abs/1703.01570}
  {arXiv:1703.01570 [hep-ph]} \BibitemShut {NoStop}%
\bibitem [{\citenamefont {Chupp}\ \emph {et~al.}(2019)\citenamefont {Chupp},
  \citenamefont {Fierlinger}, \citenamefont {Ramsey-Musolf},\ and\
  \citenamefont {Singh}}]{Chupp:2017}%
  \BibitemOpen
  \bibfield  {author} {\bibinfo {author} {\bibfnamefont {T.}~\bibnamefont
  {Chupp}}, \bibinfo {author} {\bibfnamefont {P.}~\bibnamefont {Fierlinger}},
  \bibinfo {author} {\bibfnamefont {M.}~\bibnamefont {Ramsey-Musolf}}, \ and\
  \bibinfo {author} {\bibfnamefont {J.}~\bibnamefont {Singh}},\ }\href
  {\doibase 10.1103/RevModPhys.91.015001} {\bibfield  {journal} {\bibinfo
  {journal} {Rev. Mod. Phys.}\ }\textbf {\bibinfo {volume} {91}},\ \bibinfo
  {pages} {015001} (\bibinfo {year} {2019})},\ \Eprint
  {http://arxiv.org/abs/1710.02504} {arXiv:1710.02504 [physics.atom-ph]}
  \BibitemShut {NoStop}%
\bibitem [{\citenamefont {Graner}\ \emph {et~al.}(2016)\citenamefont {Graner},
  \citenamefont {Chen}, \citenamefont {Lindahl},\ and\ \citenamefont
  {Heckel}}]{Graner:2016Hg199_EDM}%
  \BibitemOpen
  \bibfield  {author} {\bibinfo {author} {\bibfnamefont {B.}~\bibnamefont
  {Graner}}, \bibinfo {author} {\bibfnamefont {Y.}~\bibnamefont {Chen}},
  \bibinfo {author} {\bibfnamefont {E.~G.}\ \bibnamefont {Lindahl}}, \ and\
  \bibinfo {author} {\bibfnamefont {B.~R.}\ \bibnamefont {Heckel}},\ }\href
  {\doibase 10.1103/PhysRevLett.116.161601} {\bibfield  {journal} {\bibinfo
  {journal} {Phys. Rev. Lett.}\ }\textbf {\bibinfo {volume} {116}},\ \bibinfo
  {pages} {161601} (\bibinfo {year} {2016})},\ \bibinfo {note} {[Erratum:
  Phys.Rev.Lett. 119, 119901 (2017)]},\ \Eprint
  {http://arxiv.org/abs/1601.04339} {arXiv:1601.04339 [physics.atom-ph]}
  \BibitemShut {NoStop}%
\bibitem [{\citenamefont {Andreev}\ \emph {et~al.}(2018)\citenamefont {Andreev}
  \emph {et~al.}}]{ACME:2018}%
  \BibitemOpen
  \bibfield  {author} {\bibinfo {author} {\bibfnamefont {V.}~\bibnamefont
  {Andreev}} \emph {et~al.} (\bibinfo {collaboration} {ACME}),\ }\href
  {\doibase 10.1038/s41586-018-0599-8} {\bibfield  {journal} {\bibinfo
  {journal} {Nature}\ }\textbf {\bibinfo {volume} {562}},\ \bibinfo {pages}
  {355} (\bibinfo {year} {2018})}\BibitemShut {NoStop}%
\bibitem [{\citenamefont {Abel}\ \emph {et~al.}(2020)\citenamefont {Abel} \emph
  {et~al.}}]{Abel:2020}%
  \BibitemOpen
  \bibfield  {author} {\bibinfo {author} {\bibfnamefont {C.}~\bibnamefont
  {Abel}} \emph {et~al.},\ }\href {\doibase 10.1103/PhysRevLett.124.081803}
  {\bibfield  {journal} {\bibinfo  {journal} {Phys. Rev. Lett.}\ }\textbf
  {\bibinfo {volume} {124}},\ \bibinfo {pages} {081803} (\bibinfo {year}
  {2020})}\BibitemShut {NoStop}%
\bibitem [{\citenamefont {Alarcon}\ \emph {et~al.}(2022)\citenamefont {Alarcon}
  \emph {et~al.}}]{Alarcon:Snowmass2022}%
  \BibitemOpen
  \bibfield  {author} {\bibinfo {author} {\bibfnamefont {R.}~\bibnamefont
  {Alarcon}} \emph {et~al.},\ }in\ \href@noop {} {\emph {\bibinfo {booktitle}
  {{Snowmass 2021}}}}\ (\bibinfo {year} {2022})\ \Eprint
  {http://arxiv.org/abs/2203.08103} {arXiv:2203.08103 [hep-ph]} \BibitemShut
  {NoStop}%
\bibitem [{\citenamefont {Yoshinaga}\ \emph {et~al.}(2018)\citenamefont
  {Yoshinaga}, \citenamefont {Teruya}, \citenamefont {Higashiyama},\ and\
  \citenamefont {Yanase}}]{Yoshinaga:2018PTSM3}%
  \BibitemOpen
  \bibfield  {author} {\bibinfo {author} {\bibfnamefont {N.}~\bibnamefont
  {Yoshinaga}}, \bibinfo {author} {\bibfnamefont {E.}~\bibnamefont {Teruya}},
  \bibinfo {author} {\bibfnamefont {K.}~\bibnamefont {Higashiyama}}, \ and\
  \bibinfo {author} {\bibfnamefont {K.}~\bibnamefont {Yanase}},\ }\href
  {\doibase 10.7566/JPSCP.23.012034} {\bibfield  {journal} {\bibinfo  {journal}
  {JPS Conf. Proc.}\ }\textbf {\bibinfo {volume} {23}},\ \bibinfo {pages}
  {012034} (\bibinfo {year} {2018})}\BibitemShut {NoStop}%
\bibitem [{\citenamefont {Sachdeva}\ \emph {et~al.}(2019)\citenamefont
  {Sachdeva} \emph {et~al.}}]{Sachdeva:2019PRL}%
  \BibitemOpen
  \bibfield  {author} {\bibinfo {author} {\bibfnamefont {N.}~\bibnamefont
  {Sachdeva}} \emph {et~al.},\ }\href {\doibase 10.1103/PhysRevLett.123.143003}
  {\bibfield  {journal} {\bibinfo  {journal} {Phys. Rev. Lett.}\ }\textbf
  {\bibinfo {volume} {123}},\ \bibinfo {pages} {143003} (\bibinfo {year}
  {2019})},\ \Eprint {http://arxiv.org/abs/1902.02864} {arXiv:1902.02864
  [physics.atom-ph]} \BibitemShut {NoStop}%
\bibitem [{\citenamefont {Zheng}\ \emph {et~al.}(2022)\citenamefont {Zheng},
  \citenamefont {Yang}, \citenamefont {Wang}, \citenamefont {Singh},
  \citenamefont {Xiong}, \citenamefont {Xia},\ and\ \citenamefont
  {Lu}}]{Zheng:2022PRL}%
  \BibitemOpen
  \bibfield  {author} {\bibinfo {author} {\bibfnamefont {T.~A.}\ \bibnamefont
  {Zheng}}, \bibinfo {author} {\bibfnamefont {Y.~A.}\ \bibnamefont {Yang}},
  \bibinfo {author} {\bibfnamefont {S.~Z.}\ \bibnamefont {Wang}}, \bibinfo
  {author} {\bibfnamefont {J.~T.}\ \bibnamefont {Singh}}, \bibinfo {author}
  {\bibfnamefont {Z.~X.}\ \bibnamefont {Xiong}}, \bibinfo {author}
  {\bibfnamefont {T.}~\bibnamefont {Xia}}, \ and\ \bibinfo {author}
  {\bibfnamefont {Z.~T.}\ \bibnamefont {Lu}},\ }\href {\doibase
  10.1103/PhysRevLett.129.083001} {\bibfield  {journal} {\bibinfo  {journal}
  {Phys. Rev. Lett.}\ }\textbf {\bibinfo {volume} {129}},\ \bibinfo {pages}
  {083001} (\bibinfo {year} {2022})},\ \Eprint
  {http://arxiv.org/abs/2207.08140} {arXiv:2207.08140 [physics.atom-ph]}
  \BibitemShut {NoStop}%
\bibitem [{\citenamefont {Regan}\ \emph {et~al.}(2002)\citenamefont {Regan},
  \citenamefont {Commins}, \citenamefont {Schmidt},\ and\ \citenamefont
  {DeMille}}]{Regan:2002PRL}%
  \BibitemOpen
  \bibfield  {author} {\bibinfo {author} {\bibfnamefont {B.~C.}\ \bibnamefont
  {Regan}}, \bibinfo {author} {\bibfnamefont {E.~D.}\ \bibnamefont {Commins}},
  \bibinfo {author} {\bibfnamefont {C.~J.}\ \bibnamefont {Schmidt}}, \ and\
  \bibinfo {author} {\bibfnamefont {D.}~\bibnamefont {DeMille}},\ }\href
  {\doibase 10.1103/PhysRevLett.88.071805} {\bibfield  {journal} {\bibinfo
  {journal} {Phys. Rev. Lett.}\ }\textbf {\bibinfo {volume} {88}},\ \bibinfo
  {pages} {071805} (\bibinfo {year} {2002})}\BibitemShut {NoStop}%
\bibitem [{\citenamefont {Haxton}\ and\ \citenamefont
  {Henley}(1983)}]{Haxton:1983}%
  \BibitemOpen
  \bibfield  {author} {\bibinfo {author} {\bibfnamefont {W.~C.}\ \bibnamefont
  {Haxton}}\ and\ \bibinfo {author} {\bibfnamefont {E.~M.}\ \bibnamefont
  {Henley}},\ }\href {\doibase 10.1103/PhysRevLett.51.1937} {\bibfield
  {journal} {\bibinfo  {journal} {Phys. Rev. Lett.}\ }\textbf {\bibinfo
  {volume} {51}},\ \bibinfo {pages} {1937} (\bibinfo {year}
  {1983})}\BibitemShut {NoStop}%
\bibitem [{\citenamefont {Flambaum}\ \emph {et~al.}(1984)\citenamefont
  {Flambaum}, \citenamefont {Khriplovich},\ and\ \citenamefont
  {Sushkov}}]{Flambaum:1984}%
  \BibitemOpen
  \bibfield  {author} {\bibinfo {author} {\bibfnamefont {V.~V.}\ \bibnamefont
  {Flambaum}}, \bibinfo {author} {\bibfnamefont {I.~B.}\ \bibnamefont
  {Khriplovich}}, \ and\ \bibinfo {author} {\bibfnamefont {O.~P.}\ \bibnamefont
  {Sushkov}},\ }\href@noop {} {\bibfield  {journal} {\bibinfo  {journal} {Sov.
  Phys. JETP}\ }\textbf {\bibinfo {volume} {60}},\ \bibinfo {pages} {873}
  (\bibinfo {year} {1984})}\BibitemShut {NoStop}%
\bibitem [{\citenamefont {Flambaum}\ \emph {et~al.}(1986)\citenamefont
  {Flambaum}, \citenamefont {Khriplovich},\ and\ \citenamefont
  {Sushkov}}]{Flambaum:1986NPA}%
  \BibitemOpen
  \bibfield  {author} {\bibinfo {author} {\bibfnamefont {V.~V.}\ \bibnamefont
  {Flambaum}}, \bibinfo {author} {\bibfnamefont {I.~B.}\ \bibnamefont
  {Khriplovich}}, \ and\ \bibinfo {author} {\bibfnamefont {O.~P.}\ \bibnamefont
  {Sushkov}},\ }\href {\doibase 10.1016/0375-9474(86)90331-3} {\bibfield
  {journal} {\bibinfo  {journal} {Nucl. Phys. A}\ }\textbf {\bibinfo {volume}
  {449}},\ \bibinfo {pages} {750} (\bibinfo {year} {1986})}\BibitemShut
  {NoStop}%
\bibitem [{\citenamefont {Engel}(2025)}]{Engel:2025}%
  \BibitemOpen
  \bibfield  {author} {\bibinfo {author} {\bibfnamefont {J.}~\bibnamefont
  {Engel}},\ }\href {\doibase 10.1146/annurev-nucl-121423-101030} {\bibfield
  {journal} {\bibinfo  {journal} {Ann. Rev. Nucl. Part. Sci.}\ }\textbf
  {\bibinfo {volume} {75}},\ \bibinfo {pages} {129} (\bibinfo {year} {2025})},\
  \Eprint {http://arxiv.org/abs/2501.02744} {arXiv:2501.02744 [nucl-th]}
  \BibitemShut {NoStop}%
\bibitem [{\citenamefont {Parker}\ \emph {et~al.}(2015)\citenamefont {Parker}
  \emph {et~al.}}]{Parker:2015Ra225_EDM}%
  \BibitemOpen
  \bibfield  {author} {\bibinfo {author} {\bibfnamefont {R.~H.}\ \bibnamefont
  {Parker}} \emph {et~al.},\ }\href {\doibase 10.1103/PhysRevLett.114.233002}
  {\bibfield  {journal} {\bibinfo  {journal} {Phys. Rev. Lett.}\ }\textbf
  {\bibinfo {volume} {114}},\ \bibinfo {pages} {233002} (\bibinfo {year}
  {2015})},\ \Eprint {http://arxiv.org/abs/1504.07477} {arXiv:1504.07477
  [nucl-ex]} \BibitemShut {NoStop}%
\bibitem [{\citenamefont {Spevak}\ \emph {et~al.}(1997)\citenamefont {Spevak},
  \citenamefont {Auerbach},\ and\ \citenamefont {Flambaum}}]{Spevak:1996PRM}%
  \BibitemOpen
  \bibfield  {author} {\bibinfo {author} {\bibfnamefont {V.}~\bibnamefont
  {Spevak}}, \bibinfo {author} {\bibfnamefont {N.}~\bibnamefont {Auerbach}}, \
  and\ \bibinfo {author} {\bibfnamefont {V.~V.}\ \bibnamefont {Flambaum}},\
  }\href {\doibase 10.1103/PhysRevC.56.1357} {\bibfield  {journal} {\bibinfo
  {journal} {Phys. Rev. C}\ }\textbf {\bibinfo {volume} {56}},\ \bibinfo
  {pages} {1357} (\bibinfo {year} {1997})},\ \Eprint
  {http://arxiv.org/abs/nucl-th/9612044} {arXiv:nucl-th/9612044} \BibitemShut
  {NoStop}%
\bibitem [{\citenamefont {Dmitriev}\ and\ \citenamefont
  {Sen'kov}(2003)}]{Dmitriev:2003RPA2}%
  \BibitemOpen
  \bibfield  {author} {\bibinfo {author} {\bibfnamefont {V.~F.}\ \bibnamefont
  {Dmitriev}}\ and\ \bibinfo {author} {\bibfnamefont {R.~A.}\ \bibnamefont
  {Sen'kov}},\ }\href {\doibase 10.1134/1.1619505} {\bibfield  {journal}
  {\bibinfo  {journal} {Phys. Atom. Nucl.}\ }\textbf {\bibinfo {volume} {66}},\
  \bibinfo {pages} {1940} (\bibinfo {year} {2003})},\ \Eprint
  {http://arxiv.org/abs/nucl-th/0304048} {arXiv:nucl-th/0304048} \BibitemShut
  {NoStop}%
\bibitem [{\citenamefont {Dmitriev}\ \emph {et~al.}(2005)\citenamefont
  {Dmitriev}, \citenamefont {Sen'kov},\ and\ \citenamefont
  {Auerbach}}]{Dmitriev:2004PRC}%
  \BibitemOpen
  \bibfield  {author} {\bibinfo {author} {\bibfnamefont {V.~F.}\ \bibnamefont
  {Dmitriev}}, \bibinfo {author} {\bibfnamefont {R.~A.}\ \bibnamefont
  {Sen'kov}}, \ and\ \bibinfo {author} {\bibfnamefont {N.}~\bibnamefont
  {Auerbach}},\ }\href {\doibase 10.1103/PhysRevC.71.035501} {\bibfield
  {journal} {\bibinfo  {journal} {Phys. Rev. C}\ }\textbf {\bibinfo {volume}
  {71}},\ \bibinfo {pages} {035501} (\bibinfo {year} {2005})},\ \Eprint
  {http://arxiv.org/abs/nucl-th/0408065} {arXiv:nucl-th/0408065} \BibitemShut
  {NoStop}%
\bibitem [{\citenamefont {de~Jesus}\ and\ \citenamefont
  {Engel}(2005)}]{deJesus:2005SQRPA1}%
  \BibitemOpen
  \bibfield  {author} {\bibinfo {author} {\bibfnamefont {J.~H.}\ \bibnamefont
  {de~Jesus}}\ and\ \bibinfo {author} {\bibfnamefont {J.}~\bibnamefont
  {Engel}},\ }\href {\doibase 10.1103/PhysRevC.72.045503} {\bibfield  {journal}
  {\bibinfo  {journal} {Phys. Rev. C}\ }\textbf {\bibinfo {volume} {72}},\
  \bibinfo {pages} {045503} (\bibinfo {year} {2005})},\ \Eprint
  {http://arxiv.org/abs/nucl-th/0507031} {arXiv:nucl-th/0507031} \BibitemShut
  {NoStop}%
\bibitem [{\citenamefont {Dobaczewski}\ and\ \citenamefont
  {Engel}(2005)}]{Dobaczewski:2005PRL}%
  \BibitemOpen
  \bibfield  {author} {\bibinfo {author} {\bibfnamefont {J.}~\bibnamefont
  {Dobaczewski}}\ and\ \bibinfo {author} {\bibfnamefont {J.}~\bibnamefont
  {Engel}},\ }\href {\doibase 10.1103/PhysRevLett.94.232502} {\bibfield
  {journal} {\bibinfo  {journal} {Phys. Rev. Lett.}\ }\textbf {\bibinfo
  {volume} {94}},\ \bibinfo {pages} {232502} (\bibinfo {year} {2005})},\
  \Eprint {http://arxiv.org/abs/nucl-th/0503057} {arXiv:nucl-th/0503057}
  \BibitemShut {NoStop}%
\bibitem [{\citenamefont {Ban}\ \emph {et~al.}(2010)\citenamefont {Ban},
  \citenamefont {Dobaczewski}, \citenamefont {Engel},\ and\ \citenamefont
  {Shukla}}]{Ban:2010PRC}%
  \BibitemOpen
  \bibfield  {author} {\bibinfo {author} {\bibfnamefont {S.}~\bibnamefont
  {Ban}}, \bibinfo {author} {\bibfnamefont {J.}~\bibnamefont {Dobaczewski}},
  \bibinfo {author} {\bibfnamefont {J.}~\bibnamefont {Engel}}, \ and\ \bibinfo
  {author} {\bibfnamefont {A.}~\bibnamefont {Shukla}},\ }\href {\doibase
  10.1103/PhysRevC.82.015501} {\bibfield  {journal} {\bibinfo  {journal} {Phys.
  Rev. C}\ }\textbf {\bibinfo {volume} {82}},\ \bibinfo {pages} {015501}
  (\bibinfo {year} {2010})},\ \Eprint {http://arxiv.org/abs/1003.2598}
  {arXiv:1003.2598 [nucl-th]} \BibitemShut {NoStop}%
\bibitem [{\citenamefont {Teruya}\ \emph {et~al.}(2017)\citenamefont {Teruya},
  \citenamefont {Yoshinaga}, \citenamefont {Higashiyama},\ and\ \citenamefont
  {Asahi}}]{Teruya:2017PTSM1}%
  \BibitemOpen
  \bibfield  {author} {\bibinfo {author} {\bibfnamefont {E.}~\bibnamefont
  {Teruya}}, \bibinfo {author} {\bibfnamefont {N.}~\bibnamefont {Yoshinaga}},
  \bibinfo {author} {\bibfnamefont {K.}~\bibnamefont {Higashiyama}}, \ and\
  \bibinfo {author} {\bibfnamefont {K.}~\bibnamefont {Asahi}},\ }\href
  {\doibase 10.1103/PhysRevC.96.015501} {\bibfield  {journal} {\bibinfo
  {journal} {Phys. Rev. C}\ }\textbf {\bibinfo {volume} {96}},\ \bibinfo
  {pages} {015501} (\bibinfo {year} {2017})}\BibitemShut {NoStop}%
\bibitem [{\citenamefont {Yoshinaga}\ \emph {et~al.}(2020)\citenamefont
  {Yoshinaga}, \citenamefont {Yanase},\ and\ \citenamefont
  {Higashiyama}}]{Yoshinaga:2018PTSM4}%
  \BibitemOpen
  \bibfield  {author} {\bibinfo {author} {\bibfnamefont {N.}~\bibnamefont
  {Yoshinaga}}, \bibinfo {author} {\bibfnamefont {K.}~\bibnamefont {Yanase}}, \
  and\ \bibinfo {author} {\bibfnamefont {K.}~\bibnamefont {Higashiyama}},\
  }\href {\doibase 10.1088/1742-6596/1643/1/012006} {\bibfield  {journal}
  {\bibinfo  {journal} {J. Phys. Conf. Ser.}\ }\textbf {\bibinfo {volume}
  {1643}},\ \bibinfo {pages} {012006} (\bibinfo {year} {2020})}\BibitemShut
  {NoStop}%
\bibitem [{\citenamefont {Yanase}\ and\ \citenamefont
  {Shimizu}(2020)}]{Yanase:2020LSSM1}%
  \BibitemOpen
  \bibfield  {author} {\bibinfo {author} {\bibfnamefont {K.}~\bibnamefont
  {Yanase}}\ and\ \bibinfo {author} {\bibfnamefont {N.}~\bibnamefont
  {Shimizu}},\ }\href {\doibase 10.1103/PhysRevC.102.065502} {\bibfield
  {journal} {\bibinfo  {journal} {Phys. Rev. C}\ }\textbf {\bibinfo {volume}
  {102}},\ \bibinfo {pages} {065502} (\bibinfo {year} {2020})},\ \Eprint
  {http://arxiv.org/abs/2006.15142} {arXiv:2006.15142 [nucl-th]} \BibitemShut
  {NoStop}%
\bibitem [{\citenamefont {Engel}\ \emph {et~al.}(2013)\citenamefont {Engel},
  \citenamefont {Ramsey-Musolf},\ and\ \citenamefont {van
  Kolck}}]{Engel:2013review1}%
  \BibitemOpen
  \bibfield  {author} {\bibinfo {author} {\bibfnamefont {J.}~\bibnamefont
  {Engel}}, \bibinfo {author} {\bibfnamefont {M.~J.}\ \bibnamefont
  {Ramsey-Musolf}}, \ and\ \bibinfo {author} {\bibfnamefont {U.}~\bibnamefont
  {van Kolck}},\ }\href {\doibase 10.1016/j.ppnp.2013.03.003} {\bibfield
  {journal} {\bibinfo  {journal} {Prog. Part. Nucl. Phys.}\ }\textbf {\bibinfo
  {volume} {71}},\ \bibinfo {pages} {21} (\bibinfo {year} {2013})},\ \Eprint
  {http://arxiv.org/abs/1303.2371} {arXiv:1303.2371 [nucl-th]} \BibitemShut
  {NoStop}%
\bibitem [{\citenamefont {Ginges}\ and\ \citenamefont
  {Flambaum}(2004)}]{Ginges:2003review}%
  \BibitemOpen
  \bibfield  {author} {\bibinfo {author} {\bibfnamefont {J.~S.~M.}\
  \bibnamefont {Ginges}}\ and\ \bibinfo {author} {\bibfnamefont {V.~V.}\
  \bibnamefont {Flambaum}},\ }\href {\doibase 10.1016/j.physrep.2004.03.005}
  {\bibfield  {journal} {\bibinfo  {journal} {Phys. Rept.}\ }\textbf {\bibinfo
  {volume} {397}},\ \bibinfo {pages} {63} (\bibinfo {year} {2004})},\ \Eprint
  {http://arxiv.org/abs/physics/0309054} {arXiv:physics/0309054} \BibitemShut
  {NoStop}%
\bibitem [{\citenamefont {Ng}\ \emph {et~al.}(2026)\citenamefont {Ng},
  \citenamefont {Foster}, \citenamefont {Cheng}, \citenamefont {Navratil},\
  and\ \citenamefont {Malbrunot-Ettenauer}}]{Ng:2026}%
  \BibitemOpen
  \bibfield  {author} {\bibinfo {author} {\bibfnamefont {K.~B.}\ \bibnamefont
  {Ng}}, \bibinfo {author} {\bibfnamefont {S.}~\bibnamefont {Foster}}, \bibinfo
  {author} {\bibfnamefont {L.}~\bibnamefont {Cheng}}, \bibinfo {author}
  {\bibfnamefont {P.}~\bibnamefont {Navratil}}, \ and\ \bibinfo {author}
  {\bibfnamefont {S.}~\bibnamefont {Malbrunot-Ettenauer}},\ }\href {\doibase
  10.1103/m4fx-g1h6} {\bibfield  {journal} {\bibinfo  {journal} {Phys. Rev.
  Lett.}\ }\textbf {\bibinfo {volume} {136}},\ \bibinfo {pages} {112501}
  (\bibinfo {year} {2026})},\ \Eprint {http://arxiv.org/abs/2507.19811}
  {arXiv:2507.19811 [nucl-th]} \BibitemShut {NoStop}%
\bibitem [{\citenamefont {Belley}\ \emph {et~al.}(2026)\citenamefont {Belley}
  \emph {et~al.}}]{Belley:2026}%
  \BibitemOpen
  \bibfield  {author} {\bibinfo {author} {\bibfnamefont {A.}~\bibnamefont
  {Belley}} \emph {et~al.},\ }\href@noop {} {\  (\bibinfo {year} {2026})},\
  \Eprint {http://arxiv.org/abs/2605.11353} {arXiv:2605.11353 [nucl-th]}
  \BibitemShut {NoStop}%
\bibitem [{\citenamefont {Auerbach}\ \emph {et~al.}(1996)\citenamefont
  {Auerbach}, \citenamefont {Flambaum},\ and\ \citenamefont
  {Spevak}}]{Auerbach:1996}%
  \BibitemOpen
  \bibfield  {author} {\bibinfo {author} {\bibfnamefont {N.}~\bibnamefont
  {Auerbach}}, \bibinfo {author} {\bibfnamefont {V.~V.}\ \bibnamefont
  {Flambaum}}, \ and\ \bibinfo {author} {\bibfnamefont {V.}~\bibnamefont
  {Spevak}},\ }\href {\doibase 10.1103/PhysRevLett.76.4316} {\bibfield
  {journal} {\bibinfo  {journal} {Phys. Rev. Lett.}\ }\textbf {\bibinfo
  {volume} {76}},\ \bibinfo {pages} {4316} (\bibinfo {year} {1996})},\ \Eprint
  {http://arxiv.org/abs/nucl-th/9601046} {arXiv:nucl-th/9601046} \BibitemShut
  {NoStop}%
\bibitem [{\citenamefont {Flambaum}\ \emph {et~al.}(2014)\citenamefont
  {Flambaum}, \citenamefont {DeMille},\ and\ \citenamefont
  {Kozlov}}]{Flambaum:2014}%
  \BibitemOpen
  \bibfield  {author} {\bibinfo {author} {\bibfnamefont {V.~V.}\ \bibnamefont
  {Flambaum}}, \bibinfo {author} {\bibfnamefont {D.}~\bibnamefont {DeMille}}, \
  and\ \bibinfo {author} {\bibfnamefont {M.~G.}\ \bibnamefont {Kozlov}},\
  }\href {\doibase 10.1103/PhysRevLett.113.103003} {\bibfield  {journal}
  {\bibinfo  {journal} {Phys. Rev. Lett.}\ }\textbf {\bibinfo {volume} {113}},\
  \bibinfo {pages} {103003} (\bibinfo {year} {2014})},\ \Eprint
  {http://arxiv.org/abs/1406.6479} {arXiv:1406.6479 [physics.atom-ph]}
  \BibitemShut {NoStop}%
\bibitem [{\citenamefont {Flambaum}\ and\ \citenamefont
  {Feldmeier}(2020)}]{Flambaum:2020}%
  \BibitemOpen
  \bibfield  {author} {\bibinfo {author} {\bibfnamefont {V.~V.}\ \bibnamefont
  {Flambaum}}\ and\ \bibinfo {author} {\bibfnamefont {H.}~\bibnamefont
  {Feldmeier}},\ }\href {\doibase 10.1103/PhysRevC.101.015502} {\bibfield
  {journal} {\bibinfo  {journal} {Phys. Rev. C}\ }\textbf {\bibinfo {volume}
  {101}},\ \bibinfo {pages} {015502} (\bibinfo {year} {2020})},\ \Eprint
  {http://arxiv.org/abs/1907.07438} {arXiv:1907.07438 [nucl-th]} \BibitemShut
  {NoStop}%
\bibitem [{\citenamefont {Sushkov}(2024)}]{Sushkov:2024}%
  \BibitemOpen
  \bibfield  {author} {\bibinfo {author} {\bibfnamefont {O.~P.}\ \bibnamefont
  {Sushkov}},\ }\href {\doibase 10.1103/PhysRevC.110.015501} {\bibfield
  {journal} {\bibinfo  {journal} {Phys. Rev. C}\ }\textbf {\bibinfo {volume}
  {110}},\ \bibinfo {pages} {015501} (\bibinfo {year} {2024})},\ \Eprint
  {http://arxiv.org/abs/2307.04299} {arXiv:2307.04299 [nucl-th]} \BibitemShut
  {NoStop}%
\bibitem [{\citenamefont {Dobaczewski}\ \emph {et~al.}(2018)\citenamefont
  {Dobaczewski}, \citenamefont {Engel}, \citenamefont {Kortelainen},\ and\
  \citenamefont {Becker}}]{Dobaczewski:2018PRL}%
  \BibitemOpen
  \bibfield  {author} {\bibinfo {author} {\bibfnamefont {J.}~\bibnamefont
  {Dobaczewski}}, \bibinfo {author} {\bibfnamefont {J.}~\bibnamefont {Engel}},
  \bibinfo {author} {\bibfnamefont {M.}~\bibnamefont {Kortelainen}}, \ and\
  \bibinfo {author} {\bibfnamefont {P.}~\bibnamefont {Becker}},\ }\href
  {\doibase 10.1103/PhysRevLett.121.232501} {\bibfield  {journal} {\bibinfo
  {journal} {Phys. Rev. Lett.}\ }\textbf {\bibinfo {volume} {121}},\ \bibinfo
  {pages} {232501} (\bibinfo {year} {2018})},\ \Eprint
  {http://arxiv.org/abs/1807.09581} {arXiv:1807.09581 [nucl-th]} \BibitemShut
  {NoStop}%
\bibitem [{\citenamefont {Ring}\ and\ \citenamefont
  {Schuck}(1980)}]{Ring:1980}%
  \BibitemOpen
  \bibfield  {author} {\bibinfo {author} {\bibfnamefont {P.}~\bibnamefont
  {Ring}}\ and\ \bibinfo {author} {\bibfnamefont {P.}~\bibnamefont {Schuck}},\
  }\href@noop {} {\emph {\bibinfo {title} {The nuclear many-body problem}}}\
  (\bibinfo  {publisher} {Springer-Verlag},\ \bibinfo {address} {New York},\
  \bibinfo {year} {1980})\BibitemShut {NoStop}%
\bibitem [{\citenamefont {Maekawa}\ \emph {et~al.}(2011)\citenamefont
  {Maekawa}, \citenamefont {Mereghetti}, \citenamefont {de~Vries},\ and\
  \citenamefont {van Kolck}}]{Maekawa:2011}%
  \BibitemOpen
  \bibfield  {author} {\bibinfo {author} {\bibfnamefont {C.~M.}\ \bibnamefont
  {Maekawa}}, \bibinfo {author} {\bibfnamefont {E.}~\bibnamefont {Mereghetti}},
  \bibinfo {author} {\bibfnamefont {J.}~\bibnamefont {de~Vries}}, \ and\
  \bibinfo {author} {\bibfnamefont {U.}~\bibnamefont {van Kolck}},\ }\href
  {\doibase 10.1016/j.nuclphysa.2011.09.020} {\bibfield  {journal} {\bibinfo
  {journal} {Nucl. Phys. A}\ }\textbf {\bibinfo {volume} {872}},\ \bibinfo
  {pages} {117} (\bibinfo {year} {2011})},\ \Eprint
  {http://arxiv.org/abs/1106.6119} {arXiv:1106.6119 [nucl-th]} \BibitemShut
  {NoStop}%
\bibitem [{\citenamefont {Goldberger}\ and\ \citenamefont
  {Treiman}(1958)}]{Goldberger:1958}%
  \BibitemOpen
  \bibfield  {author} {\bibinfo {author} {\bibfnamefont {M.~L.}\ \bibnamefont
  {Goldberger}}\ and\ \bibinfo {author} {\bibfnamefont {S.~B.}\ \bibnamefont
  {Treiman}},\ }\href {\doibase 10.1103/PhysRev.110.1178} {\bibfield  {journal}
  {\bibinfo  {journal} {Phys. Rev.}\ }\textbf {\bibinfo {volume} {110}},\
  \bibinfo {pages} {1178} (\bibinfo {year} {1958})}\BibitemShut {NoStop}%
\bibitem [{\citenamefont {Song}\ \emph {et~al.}(2014)\citenamefont {Song},
  \citenamefont {Yao}, \citenamefont {Ring},\ and\ \citenamefont
  {Meng}}]{Song:2014}%
  \BibitemOpen
  \bibfield  {author} {\bibinfo {author} {\bibfnamefont {L.~S.}\ \bibnamefont
  {Song}}, \bibinfo {author} {\bibfnamefont {J.~M.}\ \bibnamefont {Yao}},
  \bibinfo {author} {\bibfnamefont {P.}~\bibnamefont {Ring}}, \ and\ \bibinfo
  {author} {\bibfnamefont {J.}~\bibnamefont {Meng}},\ }\href {\doibase
  https://doi.org/10.1103/PhysRevC.90.054309} {\bibfield  {journal} {\bibinfo
  {journal} {Phys. Rev. C}\ }\textbf {\bibinfo {volume} {90}},\ \bibinfo
  {pages} {054309} (\bibinfo {year} {2014})}\BibitemShut {NoStop}%
\bibitem [{\citenamefont {Zhou}\ \emph {et~al.}(2024)\citenamefont {Zhou},
  \citenamefont {Wu},\ and\ \citenamefont {Yao}}]{Zhou:2024PRC}%
  \BibitemOpen
  \bibfield  {author} {\bibinfo {author} {\bibfnamefont {E.~F.}\ \bibnamefont
  {Zhou}}, \bibinfo {author} {\bibfnamefont {X.~Y.}\ \bibnamefont {Wu}}, \ and\
  \bibinfo {author} {\bibfnamefont {J.~M.}\ \bibnamefont {Yao}},\ }\href
  {\doibase 10.1103/PhysRevC.109.034305} {\bibfield  {journal} {\bibinfo
  {journal} {Phys. Rev. C}\ }\textbf {\bibinfo {volume} {109}},\ \bibinfo
  {pages} {034305} (\bibinfo {year} {2024})},\ \Eprint
  {http://arxiv.org/abs/2311.15305} {arXiv:2311.15305 [nucl-th]} \BibitemShut
  {NoStop}%
\bibitem [{\citenamefont {Zhou}\ and\ \citenamefont
  {Yao}(2025)}]{Zhou:2025_Long}%
  \BibitemOpen
  \bibfield  {author} {\bibinfo {author} {\bibfnamefont {E.~F.}\ \bibnamefont
  {Zhou}}\ and\ \bibinfo {author} {\bibfnamefont {J.~M.}\ \bibnamefont {Yao}},\
  }\href@noop {} {\  (\bibinfo {year} {2025})},\ \Eprint
  {http://arxiv.org/abs/2511.05984} {arXiv:2511.05984 [nucl-th]} \BibitemShut
  {NoStop}%
\bibitem [{\citenamefont {Hill}\ and\ \citenamefont
  {Wheeler}(1953)}]{Hill:1953}%
  \BibitemOpen
  \bibfield  {author} {\bibinfo {author} {\bibfnamefont {D.~L.}\ \bibnamefont
  {Hill}}\ and\ \bibinfo {author} {\bibfnamefont {J.~A.}\ \bibnamefont
  {Wheeler}},\ }\href {\doibase 10.1103/PhysRev.89.1102} {\bibfield  {journal}
  {\bibinfo  {journal} {Phys. Rev.}\ }\textbf {\bibinfo {volume} {89}},\
  \bibinfo {pages} {1102} (\bibinfo {year} {1953})}\BibitemShut {NoStop}%
\bibitem [{\citenamefont {Griffin}\ and\ \citenamefont
  {Wheeler}(1957)}]{Griffin:1957}%
  \BibitemOpen
  \bibfield  {author} {\bibinfo {author} {\bibfnamefont {J.~J.}\ \bibnamefont
  {Griffin}}\ and\ \bibinfo {author} {\bibfnamefont {J.~A.}\ \bibnamefont
  {Wheeler}},\ }\href {\doibase 10.1103/PhysRev.108.311} {\bibfield  {journal}
  {\bibinfo  {journal} {Phys. Rev.}\ }\textbf {\bibinfo {volume} {108}},\
  \bibinfo {pages} {311} (\bibinfo {year} {1957})}\BibitemShut {NoStop}%
\bibitem [{\citenamefont {Meng}(2016)}]{Meng:2016Book}%
  \BibitemOpen
  \bibfield  {author} {\bibinfo {author} {\bibfnamefont {J.}~\bibnamefont
  {Meng}},\ }\href {https://www.worldscientific.com/doi/abs/10.1142/9872}
  {\emph {\bibinfo {title} {Relativistic Density Functional for Nuclear
  Structure}}},\ Vol.~\bibinfo {volume} {26}\ (\bibinfo  {publisher} {World
  Scientific},\ \bibinfo {year} {2016})\BibitemShut {NoStop}%
\bibitem [{\citenamefont {Zhao}\ \emph {et~al.}(2010)\citenamefont {Zhao},
  \citenamefont {Li}, \citenamefont {Yao},\ and\ \citenamefont
  {Meng}}]{Zhao:2010PRC}%
  \BibitemOpen
  \bibfield  {author} {\bibinfo {author} {\bibfnamefont {P.~W.}\ \bibnamefont
  {Zhao}}, \bibinfo {author} {\bibfnamefont {Z.~P.}\ \bibnamefont {Li}},
  \bibinfo {author} {\bibfnamefont {J.~M.}\ \bibnamefont {Yao}}, \ and\
  \bibinfo {author} {\bibfnamefont {J.}~\bibnamefont {Meng}},\ }\href {\doibase
  10.1103/PhysRevC.82.054319} {\bibfield  {journal} {\bibinfo  {journal} {Phys.
  Rev. C}\ }\textbf {\bibinfo {volume} {82}},\ \bibinfo {pages} {054319}
  (\bibinfo {year} {2010})}\BibitemShut {NoStop}%
\bibitem [{\citenamefont {Yao}\ \emph {et~al.}(2015)\citenamefont {Yao},
  \citenamefont {Zhou},\ and\ \citenamefont {Li}}]{Yao:2015Ra224}%
  \BibitemOpen
  \bibfield  {author} {\bibinfo {author} {\bibfnamefont {J.~M.}\ \bibnamefont
  {Yao}}, \bibinfo {author} {\bibfnamefont {E.~F.}\ \bibnamefont {Zhou}}, \
  and\ \bibinfo {author} {\bibfnamefont {Z.~P.}\ \bibnamefont {Li}},\ }\href
  {\doibase 10.1103/PhysRevC.92.041304} {\bibfield  {journal} {\bibinfo
  {journal} {Phys. Rev. C}\ }\textbf {\bibinfo {volume} {92}},\ \bibinfo
  {pages} {041304} (\bibinfo {year} {2015})},\ \Eprint
  {http://arxiv.org/abs/1507.03298} {arXiv:1507.03298 [nucl-th]} \BibitemShut
  {NoStop}%
\bibitem [{\citenamefont {Yao}\ and\ \citenamefont
  {Hagino}(2016)}]{Yao:2016Pb208}%
  \BibitemOpen
  \bibfield  {author} {\bibinfo {author} {\bibfnamefont {J.~M.}\ \bibnamefont
  {Yao}}\ and\ \bibinfo {author} {\bibfnamefont {K.}~\bibnamefont {Hagino}},\
  }\href {\doibase 10.1103/PhysRevC.94.011303} {\bibfield  {journal} {\bibinfo
  {journal} {Phys. Rev. C}\ }\textbf {\bibinfo {volume} {94}},\ \bibinfo
  {pages} {011303} (\bibinfo {year} {2016})}\BibitemShut {NoStop}%
\bibitem [{\citenamefont {Yao}\ \emph {et~al.}(2010)\citenamefont {Yao},
  \citenamefont {Meng}, \citenamefont {Ring},\ and\ \citenamefont
  {Vretenar}}]{Yao:2010}%
  \BibitemOpen
  \bibfield  {author} {\bibinfo {author} {\bibfnamefont {J.~M.}\ \bibnamefont
  {Yao}}, \bibinfo {author} {\bibfnamefont {J.}~\bibnamefont {Meng}}, \bibinfo
  {author} {\bibfnamefont {P.}~\bibnamefont {Ring}}, \ and\ \bibinfo {author}
  {\bibfnamefont {D.}~\bibnamefont {Vretenar}},\ }\href {\doibase
  10.1103/PhysRevC.81.044311} {\bibfield  {journal} {\bibinfo  {journal} {Phys.
  Rev. C}\ }\textbf {\bibinfo {volume} {81}},\ \bibinfo {pages} {044311}
  (\bibinfo {year} {2010})},\ \Eprint {http://arxiv.org/abs/0912.2650}
  {arXiv:0912.2650 [nucl-th]} \BibitemShut {NoStop}%
\bibitem [{\citenamefont {Anguiano}\ \emph {et~al.}()\citenamefont {Anguiano},
  \citenamefont {Egido},\ and\ \citenamefont {Robledo}}]{Anguiano:2001}%
  \BibitemOpen
  \bibfield  {author} {\bibinfo {author} {\bibfnamefont {M.}~\bibnamefont
  {Anguiano}}, \bibinfo {author} {\bibfnamefont {J.~L.}\ \bibnamefont {Egido}},
  \ and\ \bibinfo {author} {\bibfnamefont {L.~M.}\ \bibnamefont {Robledo}},\
  }\href {\doibase 10.1016/S0375-9474(01)01219-2} {\bibfield  {journal}
  {\bibinfo  {journal} {Nucl. Phys. A}\ }\textbf {\bibinfo {volume} {696}},\
  \bibinfo {pages} {467}},\ \Eprint {http://arxiv.org/abs/nucl-th/0105003}
  {arXiv:nucl-th/0105003} \BibitemShut {NoStop}%
\bibitem [{\citenamefont {Tajima}\ \emph {et~al.}(1992)\citenamefont {Tajima},
  \citenamefont {Flocard}, \citenamefont {Bonche}, \citenamefont
  {Dobaczewski},\ and\ \citenamefont {Heenen}}]{Tajima:1992}%
  \BibitemOpen
  \bibfield  {author} {\bibinfo {author} {\bibfnamefont {N.}~\bibnamefont
  {Tajima}}, \bibinfo {author} {\bibfnamefont {H.}~\bibnamefont {Flocard}},
  \bibinfo {author} {\bibfnamefont {P.}~\bibnamefont {Bonche}}, \bibinfo
  {author} {\bibfnamefont {J.}~\bibnamefont {Dobaczewski}}, \ and\ \bibinfo
  {author} {\bibfnamefont {P.~H.}\ \bibnamefont {Heenen}},\ }\href {\doibase
  10.1016/0375-9474(92)90101-O} {\bibfield  {journal} {\bibinfo  {journal}
  {Nucl. Phys. A}\ }\textbf {\bibinfo {volume} {542}},\ \bibinfo {pages} {355}
  (\bibinfo {year} {1992})}\BibitemShut {NoStop}%
\bibitem [{\citenamefont {Dobaczewski}\ \emph {et~al.}(2007)\citenamefont
  {Dobaczewski}, \citenamefont {Stoitsov}, \citenamefont {Nazarewicz},\ and\
  \citenamefont {Reinhard}}]{Dobaczewski:2007}%
  \BibitemOpen
  \bibfield  {author} {\bibinfo {author} {\bibfnamefont {J.}~\bibnamefont
  {Dobaczewski}}, \bibinfo {author} {\bibfnamefont {M.~V.}\ \bibnamefont
  {Stoitsov}}, \bibinfo {author} {\bibfnamefont {W.}~\bibnamefont
  {Nazarewicz}}, \ and\ \bibinfo {author} {\bibfnamefont {P.-G.}\ \bibnamefont
  {Reinhard}},\ }\href {\doibase 10.1103/PhysRevC.76.054315} {\bibfield
  {journal} {\bibinfo  {journal} {Phys. Rev. C}\ }\textbf {\bibinfo {volume}
  {76}},\ \bibinfo {pages} {054315} (\bibinfo {year} {2007})}\BibitemShut
  {NoStop}%
\bibitem [{\citenamefont {Bender}\ \emph {et~al.}(2009)\citenamefont {Bender},
  \citenamefont {Duguet},\ and\ \citenamefont {Lacroix}}]{Bender:2009}%
  \BibitemOpen
  \bibfield  {author} {\bibinfo {author} {\bibfnamefont {M.}~\bibnamefont
  {Bender}}, \bibinfo {author} {\bibfnamefont {T.}~\bibnamefont {Duguet}}, \
  and\ \bibinfo {author} {\bibfnamefont {D.}~\bibnamefont {Lacroix}},\ }\href
  {\doibase 10.1103/PhysRevC.79.044319} {\bibfield  {journal} {\bibinfo
  {journal} {Phys. Rev. C}\ }\textbf {\bibinfo {volume} {79}},\ \bibinfo
  {pages} {044319} (\bibinfo {year} {2009})}\BibitemShut {NoStop}%
\bibitem [{\citenamefont {Duguet}\ \emph {et~al.}(2009)\citenamefont {Duguet},
  \citenamefont {Bender}, \citenamefont {Bennaceur}, \citenamefont {Lacroix},\
  and\ \citenamefont {Lesinski}}]{Duguet:2009}%
  \BibitemOpen
  \bibfield  {author} {\bibinfo {author} {\bibfnamefont {T.}~\bibnamefont
  {Duguet}}, \bibinfo {author} {\bibfnamefont {M.}~\bibnamefont {Bender}},
  \bibinfo {author} {\bibfnamefont {K.}~\bibnamefont {Bennaceur}}, \bibinfo
  {author} {\bibfnamefont {D.}~\bibnamefont {Lacroix}}, \ and\ \bibinfo
  {author} {\bibfnamefont {T.}~\bibnamefont {Lesinski}},\ }\href {\doibase
  10.1103/PhysRevC.79.044320} {\bibfield  {journal} {\bibinfo  {journal} {Phys.
  Rev. C}\ }\textbf {\bibinfo {volume} {79}},\ \bibinfo {pages} {044320}
  (\bibinfo {year} {2009})}\BibitemShut {NoStop}%
\bibitem [{\citenamefont {Zhou}\ \emph {et~al.}(2025)\citenamefont {Zhou},
  \citenamefont {Wu}, \citenamefont {Xiang}, \citenamefont {Yao},\ and\
  \citenamefont {Ring}}]{Zhou:2025}%
  \BibitemOpen
  \bibfield  {author} {\bibinfo {author} {\bibfnamefont {E.~F.}\ \bibnamefont
  {Zhou}}, \bibinfo {author} {\bibfnamefont {X.~Y.}\ \bibnamefont {Wu}},
  \bibinfo {author} {\bibfnamefont {J.}~\bibnamefont {Xiang}}, \bibinfo
  {author} {\bibfnamefont {J.~M.}\ \bibnamefont {Yao}}, \ and\ \bibinfo
  {author} {\bibfnamefont {P.}~\bibnamefont {Ring}},\ }\href@noop {} {\
  (\bibinfo {year} {2025})},\ \Eprint {http://arxiv.org/abs/2504.11244}
  {arXiv:2504.11244 [nucl-th]} \BibitemShut {NoStop}%
\bibitem [{\citenamefont {{\'C}wiok}\ and\ \citenamefont
  {Nazarewicz}(1991)}]{Cwiok:1991}%
  \BibitemOpen
  \bibfield  {author} {\bibinfo {author} {\bibfnamefont {S.}~\bibnamefont
  {{\'C}wiok}}\ and\ \bibinfo {author} {\bibfnamefont {W.}~\bibnamefont
  {Nazarewicz}},\ }\href {\doibase 10.1016/0375-9474(91)90787-7} {\bibfield
  {journal} {\bibinfo  {journal} {Nucl. Phys. A}\ }\textbf {\bibinfo {volume}
  {529}},\ \bibinfo {pages} {95} (\bibinfo {year} {1991})}\BibitemShut
  {NoStop}%
\bibitem [{\citenamefont {{National Nuclear Data Center}}(2020)}]{NNDC}%
  \BibitemOpen
  \bibfield  {author} {\bibinfo {author} {\bibnamefont {{National Nuclear Data
  Center}}},\ }\href {https://www.nndc.bnl.gov/nudat2} {\enquote {\bibinfo
  {title} {{NuDat 2 Database}},}\ } (\bibinfo {year} {2020}),\ \bibinfo {note}
  {\url{https://www.nndc.bnl.gov/nudat2}}\BibitemShut {NoStop}%
\bibitem [{\citenamefont {Minkov}\ \emph {et~al.}(2024)\citenamefont {Minkov},
  \citenamefont {P{\'a}lffy}, \citenamefont {Quentin},\ and\ \citenamefont
  {Bonneau}}]{Minkov:2024}%
  \BibitemOpen
  \bibfield  {author} {\bibinfo {author} {\bibfnamefont {N.}~\bibnamefont
  {Minkov}}, \bibinfo {author} {\bibfnamefont {A.}~\bibnamefont {P{\'a}lffy}},
  \bibinfo {author} {\bibfnamefont {P.}~\bibnamefont {Quentin}}, \ and\
  \bibinfo {author} {\bibfnamefont {L.}~\bibnamefont {Bonneau}},\ }\href
  {\doibase 10.1103/PhysRevC.110.034327} {\bibfield  {journal} {\bibinfo
  {journal} {Phys. Rev. C}\ }\textbf {\bibinfo {volume} {110}},\ \bibinfo
  {pages} {034327} (\bibinfo {year} {2024})},\ \Eprint
  {http://arxiv.org/abs/2408.11010} {arXiv:2408.11010 [nucl-th]} \BibitemShut
  {NoStop}%
\bibitem [{\citenamefont {Levon}\ \emph {et~al.}(1994)\citenamefont {Levon}
  \emph {et~al.}}]{Levon:1994}%
  \BibitemOpen
  \bibfield  {author} {\bibinfo {author} {\bibfnamefont {A.~I.}\ \bibnamefont
  {Levon}} \emph {et~al.},\ }\href {\doibase 10.1016/0375-9474(94)90260-7}
  {\bibfield  {journal} {\bibinfo  {journal} {Nucl. Phys. A}\ }\textbf
  {\bibinfo {volume} {576}},\ \bibinfo {pages} {267} (\bibinfo {year}
  {1994})}\BibitemShut {NoStop}%
\bibitem [{\citenamefont {Ahmad}\ \emph {et~al.}(1982)\citenamefont {Ahmad},
  \citenamefont {Gindler}, \citenamefont {Betts}, \citenamefont {Chasman},\
  and\ \citenamefont {Friedman}}]{Ahmad:1982}%
  \BibitemOpen
  \bibfield  {author} {\bibinfo {author} {\bibfnamefont {I.}~\bibnamefont
  {Ahmad}}, \bibinfo {author} {\bibfnamefont {J.~E.}\ \bibnamefont {Gindler}},
  \bibinfo {author} {\bibfnamefont {R.~R.}\ \bibnamefont {Betts}}, \bibinfo
  {author} {\bibfnamefont {R.~R.}\ \bibnamefont {Chasman}}, \ and\ \bibinfo
  {author} {\bibfnamefont {A.~M.}\ \bibnamefont {Friedman}},\ }\href {\doibase
  10.1103/PhysRevLett.49.1758} {\bibfield  {journal} {\bibinfo  {journal}
  {Phys. Rev. Lett.}\ }\textbf {\bibinfo {volume} {49}},\ \bibinfo {pages}
  {1758} (\bibinfo {year} {1982})}\BibitemShut {NoStop}%
\bibitem [{\citenamefont {Grafen}\ \emph {et~al.}(1991)\citenamefont {Grafen}
  \emph {et~al.}}]{Grafen:1991}%
  \BibitemOpen
  \bibfield  {author} {\bibinfo {author} {\bibfnamefont {V.}~\bibnamefont
  {Grafen}} \emph {et~al.},\ }\href {\doibase 10.1103/PhysRevC.44.R1728}
  {\bibfield  {journal} {\bibinfo  {journal} {Phys. Rev. C}\ }\textbf {\bibinfo
  {volume} {44}},\ \bibinfo {pages} {R1728} (\bibinfo {year}
  {1991})}\BibitemShut {NoStop}%
\bibitem [{\citenamefont {Beck}\ \emph {et~al.}(2007)\citenamefont {Beck},
  \citenamefont {Becker}, \citenamefont {Beiersdorfer}, \citenamefont {Brown},
  \citenamefont {Moody}, \citenamefont {Wilhelmy}, \citenamefont {Porter},
  \citenamefont {Kilbourne},\ and\ \citenamefont {Kelley}}]{Beck:2007}%
  \BibitemOpen
  \bibfield  {author} {\bibinfo {author} {\bibfnamefont {B.~R.}\ \bibnamefont
  {Beck}}, \bibinfo {author} {\bibfnamefont {J.~A.}\ \bibnamefont {Becker}},
  \bibinfo {author} {\bibfnamefont {P.}~\bibnamefont {Beiersdorfer}}, \bibinfo
  {author} {\bibfnamefont {G.~V.}\ \bibnamefont {Brown}}, \bibinfo {author}
  {\bibfnamefont {K.~J.}\ \bibnamefont {Moody}}, \bibinfo {author}
  {\bibfnamefont {J.~B.}\ \bibnamefont {Wilhelmy}}, \bibinfo {author}
  {\bibfnamefont {F.~S.}\ \bibnamefont {Porter}}, \bibinfo {author}
  {\bibfnamefont {C.~A.}\ \bibnamefont {Kilbourne}}, \ and\ \bibinfo {author}
  {\bibfnamefont {R.~L.}\ \bibnamefont {Kelley}},\ }\href {\doibase
  10.1103/PhysRevLett.98.142501} {\bibfield  {journal} {\bibinfo  {journal}
  {Phys. Rev. Lett.}\ }\textbf {\bibinfo {volume} {98}},\ \bibinfo {pages}
  {142501} (\bibinfo {year} {2007})}\BibitemShut {NoStop}%
\bibitem [{\citenamefont {Seiferle}\ \emph {et~al.}(2019)\citenamefont
  {Seiferle} \emph {et~al.}}]{Seiferle:2019}%
  \BibitemOpen
  \bibfield  {author} {\bibinfo {author} {\bibfnamefont {B.}~\bibnamefont
  {Seiferle}} \emph {et~al.},\ }\href {\doibase 10.1038/s41586-019-1533-4}
  {\bibfield  {journal} {\bibinfo  {journal} {Nature}\ }\textbf {\bibinfo
  {volume} {573}},\ \bibinfo {pages} {243} (\bibinfo {year} {2019})},\ \Eprint
  {http://arxiv.org/abs/1905.06308} {arXiv:1905.06308 [nucl-ex]} \BibitemShut
  {NoStop}%
\bibitem [{\citenamefont {Yamaguchi}\ \emph {et~al.}(2019)\citenamefont
  {Yamaguchi} \emph {et~al.}}]{Yamaguchi:2019}%
  \BibitemOpen
  \bibfield  {author} {\bibinfo {author} {\bibfnamefont {A.}~\bibnamefont
  {Yamaguchi}} \emph {et~al.},\ }\href {\doibase
  10.1103/PhysRevLett.123.222501} {\bibfield  {journal} {\bibinfo  {journal}
  {Phys. Rev. Lett.}\ }\textbf {\bibinfo {volume} {123}},\ \bibinfo {pages}
  {222501} (\bibinfo {year} {2019})},\ \Eprint
  {http://arxiv.org/abs/1912.05395} {arXiv:1912.05395 [nucl-ex]} \BibitemShut
  {NoStop}%
\bibitem [{\citenamefont {Burvenich}\ \emph {et~al.}(2002)\citenamefont
  {Burvenich}, \citenamefont {Madland}, \citenamefont {Maruhn},\ and\
  \citenamefont {Reinhard}}]{Burvenich:2002PRC}%
  \BibitemOpen
  \bibfield  {author} {\bibinfo {author} {\bibfnamefont {T.}~\bibnamefont
  {Burvenich}}, \bibinfo {author} {\bibfnamefont {D.~G.}\ \bibnamefont
  {Madland}}, \bibinfo {author} {\bibfnamefont {J.~A.}\ \bibnamefont {Maruhn}},
  \ and\ \bibinfo {author} {\bibfnamefont {P.~G.}\ \bibnamefont {Reinhard}},\
  }\href {\doibase 10.1103/PhysRevC.65.044308} {\bibfield  {journal} {\bibinfo
  {journal} {Phys. Rev. C}\ }\textbf {\bibinfo {volume} {65}},\ \bibinfo
  {pages} {044308} (\bibinfo {year} {2002})},\ \Eprint
  {http://arxiv.org/abs/nucl-th/0111012} {arXiv:nucl-th/0111012} \BibitemShut
  {NoStop}%
\bibitem [{\citenamefont {Chupp}\ and\ \citenamefont
  {Ramsey-Musolf}(2015)}]{Chupp:2015}%
  \BibitemOpen
  \bibfield  {author} {\bibinfo {author} {\bibfnamefont {T.}~\bibnamefont
  {Chupp}}\ and\ \bibinfo {author} {\bibfnamefont {M.}~\bibnamefont
  {Ramsey-Musolf}},\ }\href {\doibase 10.1103/PhysRevC.91.035502} {\bibfield
  {journal} {\bibinfo  {journal} {Phys. Rev. C}\ }\textbf {\bibinfo {volume}
  {91}},\ \bibinfo {pages} {035502} (\bibinfo {year} {2015})},\ \Eprint
  {http://arxiv.org/abs/1407.1064} {arXiv:1407.1064 [hep-ph]} \BibitemShut
  {NoStop}%
\bibitem [{\citenamefont {de~Vries}\ \emph {et~al.}(2021)\citenamefont
  {de~Vries}, \citenamefont {Gnech},\ and\ \citenamefont
  {Shain}}]{deVries:2020l}%
  \BibitemOpen
  \bibfield  {author} {\bibinfo {author} {\bibfnamefont {J.}~\bibnamefont
  {de~Vries}}, \bibinfo {author} {\bibfnamefont {A.}~\bibnamefont {Gnech}}, \
  and\ \bibinfo {author} {\bibfnamefont {S.}~\bibnamefont {Shain}},\ }\href
  {\doibase 10.1103/PhysRevC.103.L012501} {\bibfield  {journal} {\bibinfo
  {journal} {Phys. Rev. C}\ }\textbf {\bibinfo {volume} {103}},\ \bibinfo
  {pages} {L012501} (\bibinfo {year} {2021})},\ \Eprint
  {http://arxiv.org/abs/2007.04927} {arXiv:2007.04927 [hep-ph]} \BibitemShut
  {NoStop}%
\bibitem [{\citenamefont {Zhang}\ \emph {et~al.}(2025)\citenamefont {Zhang},
  \citenamefont {Wang}, \citenamefont {Ding},\ and\ \citenamefont
  {Yao}}]{Zhang:2024_Letter}%
  \BibitemOpen
  \bibfield  {author} {\bibinfo {author} {\bibfnamefont {X.}~\bibnamefont
  {Zhang}}, \bibinfo {author} {\bibfnamefont {C.~C.}\ \bibnamefont {Wang}},
  \bibinfo {author} {\bibfnamefont {C.~R.}\ \bibnamefont {Ding}}, \ and\
  \bibinfo {author} {\bibfnamefont {J.~M.}\ \bibnamefont {Yao}},\ }\href
  {\doibase 10.1103/4cnl-5dnm} {\bibfield  {journal} {\bibinfo  {journal}
  {Phys. Rev. C}\ }\textbf {\bibinfo {volume} {112}},\ \bibinfo {pages}
  {L021302} (\bibinfo {year} {2025})},\ \Eprint
  {http://arxiv.org/abs/2408.00691} {arXiv:2408.00691 [nucl-th]} \BibitemShut
  {NoStop}%
\bibitem [{\citenamefont {Athanasakis-Kaklamanakis}\ \emph
  {et~al.}(2025)\citenamefont {Athanasakis-Kaklamanakis} \emph
  {et~al.}}]{Athanasakis-Kaklamanakis:2025}%
  \BibitemOpen
  \bibfield  {author} {\bibinfo {author} {\bibfnamefont {M.}~\bibnamefont
  {Athanasakis-Kaklamanakis}} \emph {et~al.},\ }\href {\doibase
  10.1038/s41586-025-09814-1} {\bibfield  {journal} {\bibinfo  {journal}
  {Nature}\ }\textbf {\bibinfo {volume} {648}},\ \bibinfo {pages} {562}
  (\bibinfo {year} {2025})},\ \Eprint {http://arxiv.org/abs/2507.05224}
  {arXiv:2507.05224 [physics.atom-ph]} \BibitemShut {NoStop}%
\end{thebibliography}

%

\end{document}